\documentclass[10pt,twocolumn,journal]{IEEEtran}
\usepackage{times}
\usepackage{amsbsy}
\usepackage{latexsym}
\usepackage{amssymb}
\usepackage{amsmath}
\usepackage[usenames]{color}
\usepackage[ansinew]{inputenc}
\usepackage[dvips ]{graphicx}
\input epsf
\usepackage{epic,eepic}
\usepackage{url}

\title{Adaptive and Iterative Multi-Branch MMSE Decision Feedback Detection Algorithms for MIMO Systems  }
\author{Rodrigo C.\ de Lamare
 \thanks{This work has been presented in part at
ICASSP 2010. R. C. de Lamare is with CETUC, Pontifical Catholic
University of Rio de Janeiro, Rua Marquês de S. Vicente, 225, Rio de
Janeiro - 22451-900, Brazil and the Communications Research Group,
Department of Electronics, University of York, York Y010 5DD, United
Kingdom.  Emails: delamare@cetuc.puc-rio.br, rcdl500@york.ac.uk} }

\date{  }
\begin{document}
 \maketitle
\begin{abstract}
{In this work, decision feedback (DF) detection algorithms based on
multiple processing branches for multi-input multi-output (MIMO)
spatial multiplexing systems are proposed. The proposed detector
employs multiple cancellation branches with receive filters that are
obtained from a common matrix inverse and achieves a performance
close to the maximum likelihood detector (MLD). Constrained minimum
mean-squared error (MMSE) receive filters designed with constraints
on the shape and magnitude of the feedback filters for the
multi-branch MMSE DF (MB-MMSE-DF) receivers are presented. An
adaptive implementation of the proposed MB-MMSE-DF detector is
developed along with a recursive least squares-type algorithm for
estimating the parameters of the receive filters when the channel is
time-varying.  A soft-output version of the MB-MMSE-DF detector is
also proposed as a component of an iterative detection and decoding
receiver structure. A computational complexity analysis shows that
the MB-MMSE-DF detector does not require a significant additional
complexity over the conventional MMSE-DF detector, whereas a
diversity analysis discusses the diversity order achieved by the
MB-MMSE-DF detector. Simulation results show that the MB-MMSE-DF
detector achieves a performance superior to existing suboptimal
detectors and close to the MLD, while requiring significantly lower
complexity.}

\end{abstract}


\section{Introduction}

\PARstart{T}{he} deployment of multiple transmit and receive
antennas in wireless communication systems can offer significant
multiplexing \cite{foschini,telatar} and diversity gains
\cite{alamouti,tarokh}. The multiplexing gains enable high spectral
efficiencies, whereas the diversity gains increase the reliability
of the links and provide low error rates. In multi-input
multi-output (MIMO) systems, the transmitter and the receiver should
be appropriately designed in order to exploit the structure of the
propagation channels.  {In a spatial multiplexing configuration, the
capacity gain grows linearly with the minimum number of transmit and
receive antennas \cite{foschini,telatar}. In this scenario, the
system can obtain substantial gains in data rate with the
transmission of individual data streams from the transmitter to the
receiver.} In order to separate these streams, a designer must
resort to MIMO detection techniques, which are similar to multiuser
detection methods \cite{verdu}. The optimal maximum likelihood (ML)
detector is too complex to be implemented in systems with a large
number of antennas. The ML solution can be alternatively computed
using sphere decoder (SD) algorithms \cite{viterbo}-\cite{shim},
which are very efficient for MIMO systems with a small number of
antennas. However, the computational complexity of SD algorithms
depends on the noise variance, the number of data streams to be
detected and the signal constellation, resulting in high
computational costs for low signal-to-noise ratios (SNR), large MIMO
systems and high-order constellations. The high computational
complexity of the ML detector and the SD algorithms in some of the
aforementioned situations have motivated the development of numerous
alternative strategies for MIMO detection. The linear detector
\cite{duel_mimo}, the successive interference cancellation (SIC)
approach used in the {Vertical-Bell Laboratories Layered Space-Time
(VBLAST)} systems \cite{vblast}-\cite{shang} and other
decision-driven detectors such as decision feedback (DF)
\cite{dhahir}-\cite{choi} are techniques that can offer attractive
trade-offs between performance and complexity. Prior work on DF
schemes has been reported with DF detectors with SIC (S-DF)
\cite{dhahir,choi,varanasi3} and DF receivers with
 {parallel interference cancellation (PIC)} (P-DF)
\cite{woodward2,delamare_mber}, combinations of these schemes
\cite{woodward2,delamare_spadf,stspadf,mdfpic} and mechanisms to
mitigate error propagation \cite{reuter,delamare_itic}. An often
criticized aspect of these sub-optimal schemes is that they
typically do not achieve the full receive-diversity order of the ML
algorithm. {This has motivated the investigation of alternative
detection strategies such as lattice-reduction (LR) schemes
\cite{windpassinger}-\cite{gan}, {QR decomposition and the
M-algorithm (QRD-M) detectors \cite{kim_qrdm,kim_qrdm2} },
probabilistic data association (PDA) \cite{jia,syang} detectors,
extensions to soft-input soft-output detectors
\cite{hochwald}-\cite{hwang}, and calls for flexible cost-effective
detection algorithms with near-ML or ML performance, which achieve
the full receive-diversity order. }

 {In this work, a DF detection strategy based on
multiple branches (MB) is proposed for MIMO systems operating in a
spatial multiplexing configuration. The proposed detection
algorithm, termed as MB-MMSE-DF and first reported in
\cite{delamare_mbdf}, employs multiple feedforward and feedback
receive filters with appropriate transformations that are obtained
from a common matrix inverse and allow the search for improved
detection candidates. To this end, the MB-MMSE-DF receiver exploits
different patterns and orderings, and selects the branch with the
highest likelihood based on an instantaneous MMSE metric.
Constrained minimum mean-squared error (MMSE) receive filters
designed with constraints on the shape and magnitude of the feedback
filters for the proposed MB-MMSE-DF receiver are devised. The
MB-MMSE-DF detector does not require a significant additional
complexity over the conventional MMSE-DF receiver since it relies on
filter realizations with different constraints on the feedback
filters, a common matrix inversion and the same second-order
statistics. An adaptive implementation of the MB-MMSE-DF detector
with a recursive least squares (RLS)-type algorithm for estimating
the parameters of the filters when the channel is time-varying is
also presented.  The optimal ordering algorithm for the MB-MMSE-DF
detector is presented along with a low-complexity suboptimal
ordering technique. A soft-input soft-output version of the
MB-MMSE-DF receiver for iterative detection and decoding using
convolutional codes is also developed. The iterative MB-MMSE-DF
receiver employs multiple detection candidates to construct a list
of log-likelihood ratios for each transmitted bit. A diversity
analysis that discusses the diversity order achieved by the
MB-MMSE-DF detector is carried out along with a computational
complexity study. The MB-MMSE-DF detector achieves a performance
close to the optimal ML detector, while it requires a reduced cost
and has a superior performance to existing sub-optimal detectors.}

The main contributions of this work are:
\\ 1) The proposal of the MB-MMSE-DF detection algorithm;
\\ 2) MMSE expressions for filter design along with shape
patterns and magnitude constraints for the
filters; \\
3) An adaptive version of the proposed detection scheme along with a
performance and a complexity analysis;
\\
 {4) An optimal ordering algorithm is presented along
with a cost-effective suboptimal ordering algorithm for the MB-MMSE-DF detector;} \\
5) An iterative MB-MMSE-DF algorithm for processing soft
estimates with convolutional codes; \\
6) An analysis of the complexity and diversity order attained by the
MB-MMSE-DF detector;
\\ 7) A comparative study of the MB-MMSE-DF and
 existing MIMO detection algorithms.

This paper is organized as follows. Section II briefly describes a
MIMO spatial multiplexing system model. Section III is devoted to
the proposed MB-MMSE-DF detection algorithm, the design of the MMSE
filters and a multistage scheme. Section IV presents the design of
the shaping matrices, the ordering and the parameter estimation
algorithms. Section V is dedicated to the development of an
iterative version of the MB-MMSE-DF detector which processes soft
information for iterative detection and decoding. Section VI
presents an analysis of the computational complexity along with the
diversity order of the MB-MMSE-DF scheme. Section VII presents and
discusses the simulation results and Section VIII draws the
conclusions.

\section{System Model}


Consider a spatial multiplexing MIMO system with $N_T$ transmit
antennas and $N_R$ receive antennas, where $N_R \geq N_T$. At each
time instant $i$, the system transmits $N_T$ symbols which are
organized into a $N_T \times 1$ vector ${\boldsymbol s}[i] = \big[
s_1[i], ~s_2[i], ~ \ldots,~ s_{N_T}[i] \big]^T$ taken from a
modulation constellation $A = \{ a_1,~a_2,~\ldots,~a_N \}$, where
$(\cdot)^T$ denotes transpose and $N=2^C$. In other words, each
symbol is carrying $C$ bits. The symbol vector ${\boldsymbol s}[i]$
is then transmitted over flat fading channels and the signals are
demodulated and sampled at the receiver, which is equipped with
$N_R$ antennas.

The received signal after demodulation, matched filtering and
sampling is organized in an $N_R \times 1$ vector ${\boldsymbol
r}[i] = \big[ r_1[i], ~r_2[i], ~ \ldots,~ r_{N_R}[i] \big]^T$ with
sufficient statistics for detection as given by
\begin{equation}
{\boldsymbol r}[i] = {\boldsymbol H} {\boldsymbol s}[i] +
{\boldsymbol n}[i],
\end{equation}
where the $N_R \times 1$ vector ${\boldsymbol n}[i]$ is a zero mean
complex circular symmetric Gaussian noise with covariance matrix
$E\big[ {\boldsymbol n}[i] {\boldsymbol n}^H[i] \big] = \sigma_n^2
{\boldsymbol I}$, where $E[ \cdot]$ stands for expected value,
$(\cdot)^H$ denotes the Hermitian operator, $\sigma_n^2$ is the
noise variance and ${\boldsymbol I}$ is the identity matrix. The
symbol vector ${\boldsymbol s}[i]$ has zero mean and a covariance
matrix $E\big[ {\boldsymbol s}[i] {\boldsymbol s}^H[i] \big] =
\sigma_s^2 {\boldsymbol I}$, where $\sigma_s^2$ is the signal power.
The elements $h_{n_R,n_T}$ of the $N_R \times N_T$ channel matrix
${\boldsymbol H}$ correspond to the complex channel gains from the
$n_T$th transmit antenna to the $n_R$th receive antenna.

\section{Multi-Branch MMSE  Decision Feedback Detection}

\begin{figure}[!htb]
\begin{center}
\def\epsfsize#1#2{1\columnwidth}
\epsfbox{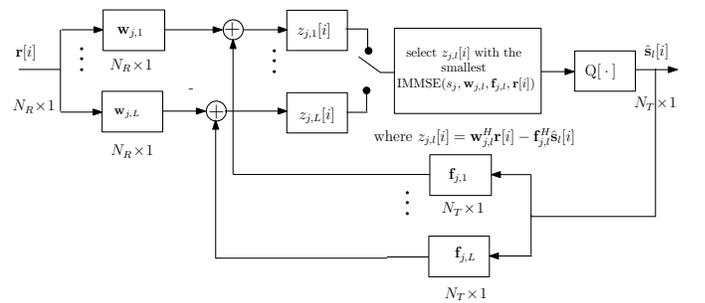} \caption{Block diagram of the proposed MB-MMSE-DF
detector and the processing of the $j$th data stream.} \label{mbdf}
\end{center}
\end{figure}

{In this section, the structure of the proposed MB-MMSE-DF detector
for MIMO systems is presented and a schematic of the detector is
shown in Fig. \ref{mbdf}. The MB-MMSE-DF detector employs multiple
pairs of MMSE receive filters in such a way that the detector can
obtain different local maxima of the likelihood function and select
the best candidate for detection according to an instantaneous MMSE
metric for each received data symbol. The receive filters are
designed based on the MMSE statistical criterion whereas the
detection and the selection of the best candidate for each received
symbol relies on an instantaneous MMSE criterion. The MB-MMSE-DF
scheme is flexible and approaches the full receive diversity
available in the system by increasing the number of branches. The
MB-MMSE-DF detector employs tasks such as MB processing, MMSE
decision feedback, and ordering that have a combined computational
cost that is substantially lower than the ML detector, which is very
simple from a mathematical  {point of view but requires} a number of
operations that is much higher than the MB-MMSE-DF and other
existing detectors.}

{In order to detect each transmitted data stream using the proposed
MB-MMSE-DF detector, the receiver linearly combines the feedforward
filter { represented by the $N_R \times 1$ vector} ${\boldsymbol
w}_{j,l}$ corresponding to the $j$-th data stream and the $l$-th
branch with the received vector ${\boldsymbol r}[i]$, subtracts the
remaining interference by linearly combining the feedback filter {
denoted by the $N_T \times 1$ vector} ${\boldsymbol f}_{j,l}$ with
the $N_T \times 1$ vector of initial decisions $\hat{\boldsymbol
s}_l[i]$ obtained from previous decisions. This process is repeated
for $L$ candidate symbols and $N_T$ data streams as described by}
\begin{equation}
\begin{split}
\label{eq:df} { z}_{j,l}[i] & = {\boldsymbol
w}^{H}_{j,l}{\boldsymbol r}[i] - {\boldsymbol f}_{j,l}^{H}
\hat{\boldsymbol s}_l[i], ~~ j  = 1, ~ \ldots, ~N_T~~ {\rm and} ~~
l=1,~\ldots,~L
\end{split}
\end{equation}
 {where the input to the decision device for the $i$th
symbol and the $j$-th stream is the $L \times 1$ vector
${\boldsymbol z}_j[i] =[z_{j,1}[i] ~\ldots ~z_{j,L}[i]]^{T}$.
 {The number of parallel branches $L$ that produce
detection candidates is a parameter that must be chosen by the
designer and is determined experimentally}. Another important design
aspect that affects the performance is the ordering algorithm which
will be discussed later on. { The goal of this work} is to employ a
reduced number of branches and yet achieve near-ML or ML
performance.}

 {The MB-MMSE-DF detector generates $L$ candidate
symbols for each data stream and then selects the best branch
according to an instantaneous MMSE metric as described by
\begin{equation}
l_{j,{\rm opt}} = \arg \min_{1 \leq l_j \leq L}   {\rm IMMSE}
(s_j[i],{\boldsymbol w}_{j,l}, {\boldsymbol f}_{j,l}, {\boldsymbol
r}[i] ), ~j=1, ~\ldots ,~ N_T \label{eq:error}
\end{equation} }
 { where
\begin{equation}{\rm IMMSE} (s_j[i],{\boldsymbol w}_{j,l},
{\boldsymbol f}_{j,l}, {\boldsymbol r}[i]) \approx |s_j[i]|^2 -
{\boldsymbol w}_{j,l}^{H} \hat{\boldsymbol R}[i] {\boldsymbol
w}_{j,l}  + {\boldsymbol f}_{j,l}^{H} \hat{\boldsymbol s}_{l}[i]
\hat{\boldsymbol s}^{H}_{l} [i]{\boldsymbol f}_{j,l}
\end{equation}
where the instantaneous MMSE metric IMMSE is produced by
 {the pair of receive filters ${\boldsymbol w}_{j,l}$
and ${\boldsymbol f}_{j,l}$}, the quantity $|s_j[i]|^2$, the
received vector ${\boldsymbol r}[i]$ and an instantaneous estimate
of the covariance matrix $\hat{\boldsymbol R}[i] = {\boldsymbol
r}[i]{\boldsymbol r}^H[i]$. Further details about the MMSE and IMMSE
expressions are included in the Appendices. }

The final detected symbol of the MB-MMSE-DF detector is obtained by
using the best branch as given by
\begin{equation}
\begin{split}
\hat{s}_{j}[i] & = Q \big[ {\boldsymbol z}_{j,l_{j,{\rm opt}}}[i]
\big]
 = Q \big[ {\boldsymbol w}^{H}_{j,l_{j,{\rm opt}}}{\boldsymbol r}[i]
- {\boldsymbol f}_{j,l_{j,{\rm opt}}}^{H} \hat{\boldsymbol
s}_{l_{j,{\rm opt}}}[i]  \big], ~j=1, ~ \ldots,~ N_T \label{eq:dec}
\end{split}
\end{equation}
where $Q( \cdot)$ is a slicing function that makes the decisions
about the symbols, which can be drawn from an M-PSK or a QAM
constellation.

\subsection{MMSE Filter Design }

 { In this part, the design of the MMSE receive
filters of the proposed MB-MMSE-DF detector is detailed by first
assuming imperfect feedback of the symbol decisions (${\boldsymbol
s} \neq \hat{\boldsymbol s}$) and then by assuming perfect feedback
(${\boldsymbol s} = \hat{\boldsymbol s}$)}. The design of the
receive filters is equivalent to determining feedforward filters
${\boldsymbol w}_{j,l}$ with $N_R$ coefficients and feedback filters
${\boldsymbol f}_{j,l}$ with $N_T$ elements subject to certain shape
constraints on ${\boldsymbol f}_{j,l}$ in accordance to the
following optimization problem
\begin{equation}
\begin{split}
\label{eq:msedfprop} {\rm min} & ~ {\rm MSE} (s_j[i],{\boldsymbol
w}_{j,l},{\boldsymbol f}_{j,l}) = E\big[ | { s}_j[i] - {\boldsymbol
w}^H_{j,l}{\boldsymbol r}[i] + {\boldsymbol
f}_{j,l}^{H}\hat{\boldsymbol s}_{l}[i] |^2 \big] \\
& {\rm subject}~{\rm to}~  {\boldsymbol S}_{j,l} {\boldsymbol
f}_{j,l} = {\boldsymbol 0} ~~ {\rm and} ~~
 ||{\boldsymbol f}_{j,l}||^2 = \gamma_{j,l} || {\boldsymbol
f}_{j,l}^c ||^2 , {\rm for} ~ j  = 1, \ldots, N_T ~{\rm and}~ l = 1,
\ldots, L,
\end{split}
\end{equation}
where the $N_T \times N_T$ shape constraint matrix is ${\boldsymbol
S}_{j,l}$, ${\boldsymbol 0}$ is a $ N_T \times 1$ constraint vector
and $\gamma_{j,l}$ is a design parameter that ranges from $0$ to $1$
and is responsible for scaling the norm of the conventional feedback
receive filter ${\boldsymbol f}_{j,l}^c$. The scaling of
${\boldsymbol f}_{j,l}^c$ results in the desired feedback receive
filter ${\boldsymbol f}_{j,l}$. The expectation operator is taken
over the random parameters ${\boldsymbol s}[i]$ and ${\boldsymbol
r}[i]$ assuming that ${\boldsymbol n}[i]$ and ${\boldsymbol s}[i]$
are statistically independent, and that the entries of ${\boldsymbol
s}[i]$ and ${\boldsymbol n}[i]$ are independent and identically
distributed random variables.  {The role of the shape constraint
matrix ${\boldsymbol S}_{j,l}$ is to choose the feedback connections
which will be used in the interference cancellation. If a designer
employs multiple branches and shape constraint matrices along with
different orderings then multiple candidates for detection can be
generated, resulting in an improved receiver performance. The
rationale for scaling the norm of the feedback filter is to reduce
the impact of the error propagation and improve the performance of
the receiver. This is accomplished by judiciously adjusting the
scaling of the norm and employing the value which minimizes the
error propagation}.

 {In what follows, the optimal MMSE receive filters
based on the proposed optimization in (\ref{eq:msedfprop}) are
derived.} By resorting to the method of Lagrange multipliers,
computing the gradient vectors of the Lagrangian with respect to
${\boldsymbol w}_{j,l}$ and ${\boldsymbol f}_{j,l}$, equating them
to null vectors and rearranging the terms, we obtain ${\rm for} ~ j
= 1, \ldots, N_T$ and $l = 1, \ldots, L$
\begin{equation}
\label{eq:dfeprop1} {\boldsymbol w}_{j,l}^{\rm MMSE} = {\boldsymbol
R}^{-1}({\boldsymbol p}_{j} + {\boldsymbol Q}{\boldsymbol f}_{j,l}),
\end{equation}  {
\begin{equation}
\begin{split}
\label{eq:dfeprop2} {\boldsymbol f}_{j,l}^{\rm MMSE} & =
\frac{\beta_{j,l}}{\sigma_s^2} {\boldsymbol \Pi}_{j,l} ({\boldsymbol
Q}^H{\boldsymbol w}_{j,l} - {\boldsymbol t}_{j}),
\end{split}
\end{equation} }
where
\begin{equation}
{\boldsymbol \Pi}_{j,l} = {\boldsymbol I} - {\boldsymbol
S}_{j,l}^H({\boldsymbol S}_{j,l}^H {\boldsymbol
S}_{j,l})^{-1}{\boldsymbol S}_{j,l}
\end{equation}
{is a projection matrix that ensures the shape constraint
${\boldsymbol S}_{j,l}$ on the feedback filter, $\beta_{j,l} = (1 -
\mu_{j,l})^{-1}$ is the parameter that controls the ability
 {of} the MB-MMSE-DF detector to mitigate error
propagation with values $0 \leq \beta_{j,l} \leq 1$, and $\mu_{j,l}$
is the Lagrange multiplier.  {It should be remarked that the inverse
$({\boldsymbol S}_{j,l}^H {\boldsymbol S}_{j,l})^{-1}$ might not
exist. In these situations, a pseudo-inverse is computed.} The
relationship between $\beta_{j,l}$ and $\gamma_{j,l}$ is not in
closed-form except for the extreme values when we have $\beta_{j,l}
=0$ and $\beta_{j,l} =1 $ for $\gamma_{j,l}=0$ (standard linear MMSE
detector) and $\gamma_{j,l}=1$ ( standard MB-MMSE-DF detector),
respectively.  {The optimization of the parameter $\beta_{j,l}$ has
been done with the aid of simulations because there is no
closed-form solution to obtain $\beta_{j,l}$. The simulation
approach has indicated that the performance is improved for a range
of parameters between $0.6$ and $0.7$.} This range of parameters was
verified to consistently produce good results for all the scenarios
investigated with the MB-MMSE-DF detector.} The $N_R \times N_R$
covariance matrix of the input data vector is ${\boldsymbol
R}=E[{\boldsymbol r}[i]{\boldsymbol r}^H[i]]$, ${\boldsymbol p}_j =
E[{\boldsymbol r}[i] s_j^*[i]]$, ${\boldsymbol Q} = E\big[
{\boldsymbol r}[i] \hat{\boldsymbol s}_{l}^{H}[i] \big]$, and
${\boldsymbol t}_j = E[\hat{\boldsymbol s}_{l}[i] {s}_j^*[i]]$ is
the $N_T \times 1$ vector of correlations between $\hat{\boldsymbol
s}_{l}[i]$ and ${s}_j^*[i]$.  {Substituting (\ref{eq:dfeprop2}) into
(\ref{eq:dfeprop1}) and then further manipulating the expressions we
arrive at the following MMSE receive filter expressions}
\begin{equation}
\label{eq:dfeprope1} {\boldsymbol w}_{j,l}^{\rm MMSE} = \big(
{\boldsymbol R} - \beta_{j,l}{\boldsymbol Q} {\boldsymbol \Pi}_{j,l}
{\boldsymbol Q}^H \big)^{-1} \big({\boldsymbol p}_{j} -
\beta_{j,l}{\boldsymbol \Pi}_{j,l} {\boldsymbol t}_j  \big),
\end{equation}
\begin{equation}
\begin{split}
\label{eq:dfeprope2} {\boldsymbol f}_{j,l}^{\rm MMSE} & =
\frac{\beta_{j,l}}{\sigma_s^2} {\boldsymbol \Pi}_{j,l} \big(
{\boldsymbol Q}^H \big( {\boldsymbol R} - \beta_{j,l}{\boldsymbol Q}
{\boldsymbol \Pi}_{j,l} {\boldsymbol Q}^H \big)^{-1}
\big({\boldsymbol p}_{j} - \beta_{j,l}{\boldsymbol Q}{\boldsymbol
\Pi}_{j,l} {\boldsymbol t}_j \big)
 - {\boldsymbol t}_{j}).
\end{split}
\end{equation}
The above expressions only depend on statistical quantities, and
consequently on the channel matrix ${\boldsymbol H}$, the symbol and
noise variances $\sigma^2_s$ and $\sigma^2_n$, respectively, and the
constraints.  However, the matrix inversion required for computing
${\boldsymbol w}_{j,l}$ is different for each branch and data
stream, thereby rendering the scheme computationally less efficient.
{The expressions obtained in (\ref{eq:dfeprop1}) and
(\ref{eq:dfeprop2}) are equivalent to those in (\ref{eq:dfeprope1})
and (\ref{eq:dfeprope2}), and only require iterations between them
for an equivalent performance. }  { A key advantage of using
(\ref{eq:dfeprop1}) and (\ref{eq:dfeprop2}) is that they only
require a single matrix inversion that is common to all branches and
two iterations prior to their use, whereas in (\ref{eq:dfeprope1})
and (\ref{eq:dfeprope2}) there is a matrix inversion associated with
each branch. For this reason, in what follows the expressions in
(\ref{eq:dfeprop1}) and (\ref{eq:dfeprop2}) are adopted and further
simplified.}

 {As briefly explained above, the expressions in
(\ref{eq:dfeprop1}) and (\ref{eq:dfeprop2}) can be simplified by
evaluating the expected values. By using the fact that ${\boldsymbol
t}_j= {\boldsymbol 0}$ for interference cancellation as
$\hat{\boldsymbol s}_{l}[i]$ does not contain $s_{j}$ and assuming
perfect feedback (${\boldsymbol s} = \hat{\boldsymbol s}$), the
following expressions are obtained}
\begin{equation}
\label{eq:dfe1} {\boldsymbol w}_{j,l}^{\rm MMSE} = \big(
{\boldsymbol H}{\boldsymbol H}^H + {\sigma_n^2}/{\sigma_s^2 }
{\boldsymbol I} \big)^{-1} {\boldsymbol H} ({\boldsymbol \delta}_j +
{\boldsymbol f}_{j,l}),
\end{equation}
\begin{equation}
\label{eq:dfe2} {\boldsymbol f}_{j,l}^{\rm MMSE} = \beta_{j,l}
{\boldsymbol \Pi}_{j,l} {\boldsymbol H}^H {\boldsymbol w}_{j,l}  ,
\end{equation}
where ${\boldsymbol \delta}_j = [ \underbrace{0 \ldots 0}_{j-1} ~1~
\underbrace{0 \ldots 0}_{N_T-j-2}]^T$ is a $N_T \times 1$ vector
with a one in the $j$th element and zeros elsewhere. A step-by-step
derivation of the filters is shown in Appendix I. The proposed
MB-MMSE-DF detector expressions above require the channel matrix
${\boldsymbol H}$ (in practice an estimate of it) and the noise
variance $\sigma_n^2$ at the receiver. In terms of complexity, it
requires for each branch $l$ the inversion of an $N_R \times N_R$
matrix and other operations with complexity $O(N_R^3)$. However, the
expressions obtained in (\ref{eq:dfeprop1}) and (\ref{eq:dfeprop2})
for the general case, and in (\ref{eq:dfe1}) and (\ref{eq:dfe2}) for
the case of perfect feedback, reveal that the most expensive
operations, i.e., the matrix inversions, are identical for all
branches. Therefore, the design of receive filters for the multiple
branches only requires further additions and multiplications of the
matrices. Moreover, it can be verified that the filters
${\boldsymbol w}_{j,l}^{\rm MMSE}$ and ${\boldsymbol f}_{j,l}^{\rm
MMSE}$ are dependent on one another, which means the designer has to
iterate them before applying the detector.  {It has been verified by
simulations with different system parameters and by comparing the
resulting parameters of the receive filters with those obtained by
(\ref{eq:dfeprope1}) and (\ref{eq:dfeprope2}) that it suffices to
employ two iterations of (\ref{eq:dfe1}) and (\ref{eq:dfe2}) to have
a performance equivalent to that obtained by using
(\ref{eq:dfeprope1}) and (\ref{eq:dfeprope2}). For this reason, we
employ the receive filters of (\ref{eq:dfe1}) and (\ref{eq:dfe2})
with two iterations in the proposed MB-MMSE-DF detector.}

The MMSE associated with the filters ${\boldsymbol w}_{j,l}^{\rm
MMSE}$ and ${\boldsymbol f}_{j,l}^{\rm MMSE}$ and the statistics of
the data symbols $s_j[i]$ is given by  {
\begin{equation}
\begin{split}
\underbrace{{\rm MMSE} (s_j[i],{\boldsymbol w}_{j,l}^{\rm MMSE},
{\boldsymbol f}_{j,l}^{\rm MMSE})}_{{\rm MMSE}_j}  & = \sigma_s^2 -
{\boldsymbol w}_{j,l}^{H, ~{\rm MMSE}} {\boldsymbol R} {\boldsymbol
w}_{j,l}^{\rm MMSE}  + {\boldsymbol f}_{j,l}^{H, ~{\rm MMSE}}
{\boldsymbol f}_{j,l}^{\rm MMSE}, 
\end{split}
\end{equation}}
where $\sigma_s^2 = E[|s_j[i]|^2]$ is the variance of the desired
symbol. A detailed derivation of the MMSE associated with the
receive filters is shown in Appendix II along with connections with
the MMSE achieved by conventional DF detectors.

\subsection{Multi-stage Detection for the MB-MMSE-DF}

In this subsection, algorithms for error propagation mitigation are
presented and incorporated into the structure of the MB-MMSE-DF
detection scheme. The strategy is based on iterative multi-stage
detection \cite{woodward2,delamare_spadf} that gradually refines the
decision vector and improves the overall performance. { It is}
incorporated into the MB-MMSE-DF scheme and the improvements in the
detection performance are then investigated.

\begin{figure}[!htb]
\begin{center}
\def\epsfsize#1#2{1\columnwidth}
\epsfbox{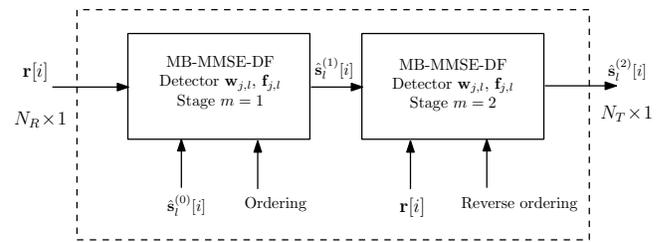} \caption{Block diagram of the proposed two-stage
MB-MMSE-DF detector.} \label{ms_mbdf}
\end{center}
\end{figure}

The basic principle underlying multi-stage detection is to
iteratively refine the estimates of the decision vector used in DF
receivers \cite{woodward2,delamare_spadf} and mitigate error
propagation. An advantage of multi-stage detection that has not been
exploited for the design of MIMO detectors is that of equalizing the
performance of the detectors over the data streams. Since V-BLAST or
DF detection usually favors certain data streams (the last detected
ones) with respect to performance, this might be important for some
applications where fairness or uniform performance is required
between the data streams. {This concept is incorporated into the
proposed MB-MMSE-DF scheme and the MMSE design of the previous
subsection. An MB-MMSE-DF scheme is employed in each stage and the
estimates of the decision vector are gradually refined as
illustrated in Fig. \ref{ms_mbdf}.} Specifically, a multi-stage
algorithm for the MB-MMSE-DF can be described by
\begin{equation}
{z}^{(m+1)}_{j,l}(i)= {\boldsymbol w}_{j,l}^{H, ~\rm
MMSE}{\boldsymbol r}[i] - {\boldsymbol f}_{j,l}^{H,~ {\rm MMSE}}
\hat{\boldsymbol s}^{(m)}_{l}[i], ~~ m=0,~1,~\ldots,~M,
\end{equation}
where the MMSE filters ${\boldsymbol w}_{j,l}^{\rm MMSE}$ and
${\boldsymbol f}_{j,l}^{\rm MMSE}$ are designed with the approach
detailed in the previous subsection, $M$ denotes the number of
stages and $\hat{\boldsymbol s}^{(m)}_{l}[i]$ is the vector of
tentative decisions from the preceding iteration that is described
by  {
\begin{equation}
\hat{ s}^{(0)}_{j,l}[i] = {\rm Q} \Big( {\boldsymbol w}_{j,l}^{H,
~\rm MMSE}{\boldsymbol r}[i] \Big), ~~ j,l = 1,~\ldots, ~N_T,
\end{equation}
\begin{equation}
\hat{ s}^{(m)}_{j,l_{j,{\rm opt}}}[i] = {\rm Q} \Big(
z^{(m)}_{j,l_{j,{\rm opt}}}[i] \Big), ~ m = 1, ~\ldots, ~M,
\end{equation}
where the number of stages $M$ depends on the scenario.}

In order to equalize the performance over the data streams, an
M-stage structure is considered. The first stage is
 {an} MB-MMSE-DF scheme with filters ${\boldsymbol
w}_{j,l}^{\rm MMSE}$ and ${\boldsymbol f}_{j,l}^{\rm MMSE}$. The
tentative decisions are passed to the second stage, which consists
of another MB-MMSE-DF scheme with {  the same receive filters} that
uses the decisions of the first stage and so successively. The
resulting multi-stage MIMO detection scheme is denoted I-MB-MMSE-DF.
The output of the second stage of the resulting scheme is
\begin{equation}
z_{j,l}^{(2)}[i]=[{\boldsymbol T}{\boldsymbol w}_{j,l}^{\rm
MMSE}]^{H}{\boldsymbol r}[i] - [{\boldsymbol T}{\boldsymbol
f}_{j,l}^{\rm MMSE}]^{H} \hat{\boldsymbol s}^{(1)}_{l_{j,{\rm
opt}}}[i],
\end{equation}
 {where $z_{j,l}^{(2)}[i]$ is the output of $j$th data
stream after multi-stage detection with $M=2$ stages}, ${\boldsymbol
T}$ is a square permutation matrix with ones along the reverse
diagonal and zeros elsewhere. When multiple stages are used, it is
beneficial to demodulate the data streams successively and in
reverse order relative to the first branch of the MB-MMSE-DF
detector. The role of reversing the cancellation order in successive
stages is to equalize the performance of the users over the
population or at least reduce the performance disparities. It
provides a better performance than keeping the same ordering as the
last decoded users in the first stage tend to be favored by the
reduced interference. The rationale is that the performance can be
improved by using the data streams that benefited from interference
cancellation (last decoded ones) as the first ones to be decoded in
the second stage. Additional stages can be included, although the
results suggest that the gains in performance are marginal for more
than two stages. Hence, the two-stage scheme is adopted for the rest
of this work.

\section{Design of Cancellation Patterns, Ordering and Adaptive Algorithms}

 {In this section, the design of the shape constraint
matrices ${\boldsymbol S}_{j,l}$ is detailed and their choices are
motivated. An optimal and a suboptimal ordering algorithms are
described for the interference cancellation. An adaptive version of
the MB-MMSE-DF detector with RLS-type algorithms is also devised.}

\subsection{Design of Cancellation Patterns}

 {The idea of the shape constraint matrices
${\boldsymbol S}_{j,l}$ is to modify the structure of the feedback
filters ${\boldsymbol f}_{j,l}$ in such a way that only the selected
feedback elements of ${\boldsymbol f}_{j,l}$ will be used to cancel
the interference between the data streams. The feedback connections
perform interference cancellation with a chosen ordering. If a
designer employs multiple branches and shape constraint matrices
along with different orderings then multiple candidates for
detection can be generated, resulting in an improved receiver
performance. The matrices ${\boldsymbol S}_{j,l}$ for the $N_T$ data
streams and for the $L$ branches of the MB-MMSE-DF detector can be
stored at the receiver and used either online or offline in the
design of the feedback filters ${\boldsymbol f}_{j,l}$.} In
particular, with this approach the ML solution can be searched from
different points of the likelihood function using an MMSE-type
detector as the starting point. Specifically, the aim is to design
and shape the filters ${\boldsymbol f}_{j,l}$ for the $N_T$ data
streams and the $L$ branches with the $N_T \times N_T$ matrices
${\boldsymbol S}_{j,l}$ such that constraint vector is a null
vector. This corresponds to allowing feedback connections of only a
subgroup of data streams. For the first branch of detection ($l=1$),
the successive cancellation used in the VBLAST \cite{vblast} can be
employed which corresponds mathematically to
\begin{equation}
\begin{split}
{\boldsymbol S}_{j,l} {\boldsymbol f}_{j,l} & = {\boldsymbol 0}, ~~
l=1 \\
{\boldsymbol S}_{j,l} & = \left[ \begin{array}{cc} {\boldsymbol
0}_{N_{T}-j+1,N_{T}-j+1} & {\boldsymbol 0}_{ N_{T}-j+1, j-1} \\
{\boldsymbol 0}_{j-1, N_{T}-j+1} & {\boldsymbol I}_{j-1,j-1}
\end{array} \right], ~~ j=1, \ldots, N_T,
\end{split}
\end{equation}
where ${\boldsymbol 0}_{m,n}$ denotes an $m \times n$-dimensional
matrix full of zeros, and ${\boldsymbol I}_m$ denotes an
$m$-dimensional identity matrix.  {Interestingly, when detecting a
data stream of interest the feedback connection associated with it
cannot be used to subtract interference because it will simply
cancel the data stream of interest itself. This is well known in the
literature of decision feedback receivers
\cite{woodward2,delamare_spadf} and is the reason for using these
structures with constraints.} For the remaining branches, an
approach based on permutations of the structure of the matrices
${\boldsymbol S}_{j,l}$ is adopted, which is given by
\begin{equation}
\begin{split}
{\boldsymbol S}_{j,l} {\boldsymbol f}_{j,l} & = {\boldsymbol 0}, ~~
l=2, \ldots, L \\
{\boldsymbol S}_{j,l} & = \phi_l \left[ \begin{array}{cc}
{\boldsymbol
0}_{N_{T}-j+1,N_{T}-j+1} & {\boldsymbol 0}_{ N_{T}-j+1, j-1} \\
{\boldsymbol 0}_{j-1, N_{T}-j+1} & {\boldsymbol I}_{j-1,j-1}
\end{array} \right], ~~
j=1, \ldots, N_T,
\end{split}
\end{equation}
 {where the operator $\phi_l[ \cdot ]$ permutes the
elements of the argument matrix such that this results in different
cancellation patterns. For instance, the non-zero elements of the
feedback filter ${\boldsymbol f}_{j,l}$ are chosen according to the
shape constraint matrices. The permutations for the different
branches will change the non-zero elements and allow the receiver to
obtain detection candidates from different interference cancellation
patterns. Although the above structure is imposed to determine the
number of feedback connections for each data stream, it might result
in a projection matrix ${\boldsymbol \Pi}_{j,l}$ whose inverse
$({\boldsymbol S}_{j,l}^H {\boldsymbol S}_{j,l})^{-1}$ does not
exist. In these situations, a pseudo-inverse is computed.
 {While there are different permutations or rotations,
the permutations employed are straightforward to implement and will
simply change the positions of the non-zero coefficients of the
feedback filters. Specifically, the permutation implemented by the
function $\phi_l[ \cdot ]$ is employed together
with the ordering to generate a list of candidates for detection.} 
The MB-MMSE detector then chooses a candidate out of $L$ branches
for each data stream which benefits from the interference
cancellation, thereby processing a data stream that is free or has a
reduced level of interference. This increases the diversity order of
the MB-MMSE-DF detector, as will be explained in the analysis of the
MB-MMSE-DF detector.}

An alternative approach for shaping the constraint matrices
${\boldsymbol S}_{j,l}$ for one of the $L$ branches is to use a
parallel interference cancellation (PIC) approach \cite{woodward2}
and design the matrices as follows
\begin{equation}
\begin{split}
{\boldsymbol S}_{j,l} {\boldsymbol f}_{j,l} & = {\boldsymbol 0}, ~~
l = ~1,~ 2,~ \ldots,~ L, \\
{\boldsymbol S}_{j,l} & =  {\boldsymbol I}_{N_{T}}  - {\rm diag}
~({\boldsymbol \delta}_j), ~~ j=1, \ldots, N_T,
\end{split}
\end{equation}
 {where ${\boldsymbol \delta}_j$ is an $N_T \times 1$
vector with a one in the $j$-th position and zeros elsewhere. The
PIC requires an initial vector of decisions obtained with the
feedforward filters ${\boldsymbol w}_{j,l}$. A problem with the PIC
approach of \cite{woodward2} is that it is prone to error
propagation due to the cancellation of all but the stream of
interest.}

\subsection{Ordering Algorithms}

 {The aim of an ordering algorithm in a MIMO system is
to obtain a sequence for interference cancellation that optimizes a
given criterion. For a conventional SIC detector, the optimal
ordering algorithm must test $N_T!$ possibilities with the objective
of minimizing the BER \cite{varanasi3,hughes}. A common alternative
to this exhaustive search is to employ a technique based on the norm
of the channels, the MMSE or the signal-to-interference-plus-noise
ratio (SINR) \cite{foschini,vblast}. The proposed MB-MMSE-DF
detector operates with an ordering based on the MMSE and the goal is
to find the best performing set of orderings over $L$ branches. The
optimal ordering algorithm for the MB-MMSE-DF detector with $L>1$,
which minimizes the MMSE for each data stream, requires testing
$\underbrace{N_T!. N_T! \ldots N_T!}_{L}$ possibilities and is given
by
\begin{equation}
\begin{split}
\{ o_{1,l}, \ldots, o_{N_T,l} \}_{\rm opt} & = \arg \min_{ o_{1,l},
\ldots, o_{N_T,l}} \sum_{l=1}^{L} \sum_{j=1}^{N_T} {\rm MMSE}_j,
{\rm for} ~ l  = 1,~ \ldots, ~L. \label{opt_ord}
\end{split}
\end{equation}}
{The rationale for this algorithm is to find the optimal ordering
for each branch, which employs the MMSE over the $L$ branches to
find the best performing set of orderings. Again, this requires
testing $\underbrace{N_T!. N_T! \ldots N_T!}_{L}$
possibilities. 
The computational complexity of the algorithm in (\ref{opt_ord}) can
increase significantly for large $N_T$ and $L$. For this reason, a
suboptimal ordering algorithm is also presented for the MB-MMSE-DF
detector.}

{In the proposed suboptimal ordering algorithm, a simplified
strategy is presented based on the maximization of the difference
between the MMSE values obtained for each data stream. For the first
branch, an ordering algorithm based on increasing values of the MMSE
is considered, and this is equivalent to an ordering according to
the maximization of the SINR for a single branch. The ordering of
the remaining branches (for the case with $L>1$) depends on the
maximization of the difference between the MMSE of different data
streams and is given by
\begin{equation}
\begin{split}
o_{j,l} & = \arg \max_{n}  \sum_{q=1}^{j-1}|{\rm MMSE}_{n} - {\rm
MMSE}_{o_{j,q}} |, ~~ {\rm for} ~ l = 2,~ \ldots, ~L ~{\rm and}~
j,n=1, \ldots, N_T \\ ~~ & {\rm subject ~to} ~~  {\rm
MMSE}_{o_{j,l}} \neq {\rm MMSE}_{o_{q,l}},~ q =1,\ldots, j-1.
\label{sub_ord}
\end{split}
\end{equation}}
{The principle behind the ordering given by (\ref{sub_ord}) with the
multiple branches is to benefit a given data stream or group for
each decoding branch. Following this approach, a data stream that
for a given ordering appears to be in an unfavorable scenario (with
more interference) can benefit in other parallel branches by being
detected in a situation with less interference, increasing the
diversity order of the MB-MMSE-DF detector. In other words, the
algorithm attempts to obtain orderings that are associated with the
largest Euclidean distance between values of MMSE for each data
stream as illustrated in Table I. This heuristic turns out to work
very well as it will be shown later.}  {The ordering algorithm in
(\ref{sub_ord}) requires a number of operations (subtractions,
modulus, and comparisons) that are linear in the number of data
streams ($N_T$) and branches ($L$), i.e., $O(N_TL)$.} {In the case
of static channels, the ordering algorithms can be employed only
once at the beginning of the transmission. In the case of
time-varying channels and whenever the MMSE obtained changes, the
ordering algorithms need to be re-computed in order to ensure an
optimized performance. }

\begin{table}[h]
\centering%
\caption{ Proposed suboptimal ordering algorithm.}
\begin{tabular}{l}
\hline  1. Ordering $O_l = \{o_{1,l} \ldots o_{j,l} \ldots o_{N_T,l} \}$ for branch $l=1$:  \\
~~~ Compute MMSE for each stream: ${\rm MMSE}_j = \sigma_j -
{\boldsymbol h}_j^H{\boldsymbol R}^{-1}{\boldsymbol h}_j$ \\
~~~ Calculate the ordering based on increasing values of ${\rm MMSE}_j$: \\
~~~~~~$o_{j,l} = \arg \min_{j} {\rm MMSE}_j$, for $j=1, \ldots, N_T$ \\
~~~~~~$ {\rm subject~
to}~ {\rm MMSE}_j \geq {\rm MMSE}_{o_{q,l}},~ q =1,\ldots, j-1 $\\

2. Ordering $O_l = \{o_{1,l} \ldots o_{j,l} \ldots o_{N_T,l} \}$ for branches $l=2, \ldots, L$: \\
~~~~~~$o_{j,l} = \arg \max_{n}  \sum_{q=1}^{j-1}|{\rm MMSE}_{n} - {\rm MMSE}_{o_{j,q}} | $, for $j,n=1, \ldots, N_T$ \\
~~~~~~$ {\rm subject~
to}~ {\rm MMSE}_{o_{j,l}} \neq {\rm MMSE}_{o_{q,l}},~ q =1,\ldots, j-1  $\\
 \hline
\end{tabular}
\end{table}



\subsection{Adaptive MB-MMSE-DF with RLS Algorithms}

In this part, an adaptive version of the MB-MMSE-DF detector with an
RLS-type algorithm is developed. The aim is to reduce the required
computational complexity of the expressions in (\ref{eq:dfeprop1})
and (\ref{eq:dfeprop2}) from $O(N_R^3)$ to $O(N_R^2)$, and equip the
proposed MB-MMSE-DF detector with the ability to track time-varying
channels. The procedure to estimate ${\boldsymbol R}^{-1}$ employs
the matrix inversion lemma \cite{haykin}:
\begin{equation}
{\boldsymbol k}[i] = \frac{\lambda^{-1} \hat{\boldsymbol R}[i-1]
{\boldsymbol r}[i] }{1+ \lambda^{-1} {\boldsymbol r}^H[i]
\hat{\boldsymbol R}[i-1] {\boldsymbol r}[i]}, \label{gain}
\end{equation}
\begin{equation}
\hat{\boldsymbol R}[i] = \lambda^{-1} \hat{\boldsymbol R}[i-1] -
\lambda^{-1} {\boldsymbol k}[i] {\boldsymbol r}^H[i] {\boldsymbol
R}^{-1}[i-1]
\end{equation}
where $0\ll \lambda<1$ is a forgetting factor that is chosen
according to the environment.  {The estimates of ${\boldsymbol
p}_k[i]$ and ${\boldsymbol Q}[i]$ are then computed with the
following recursions}
\begin{equation}
\hat{\boldsymbol Q}[i] = \lambda \hat{\boldsymbol Q}[i-1] +
{\boldsymbol r}[i] \hat{\boldsymbol s}_{l_{\rm opt}}^H[i],
\end{equation}
\begin{equation}
\hat{\boldsymbol p}_j[i] = \lambda \hat{\boldsymbol p}[i-1] +
{\boldsymbol r}[i]s_{j,l_{\rm opt}}^*[i], l =1, ~\ldots, L
\end{equation}
where the decision vector $\hat{\boldsymbol s}_{l_{\rm opt}}[i] =
Q(\hat{\boldsymbol w}^{H}_{j,l}[i-1]{\boldsymbol r}[i] -
\hat{\boldsymbol f}_{j,l}^{H}[i-1] \hat{\boldsymbol s}_{l}[i]$ is
obtained with the filters of the previous time instant. The
feedforward filters for $l = 1, ~ \ldots, L$ and $j= 1,~\ldots, N_T$
are computed by
\begin{equation}
\hat{\boldsymbol w}_{j,l}[i] = \hat{\boldsymbol R}^{-1}[i]
 (\hat{\boldsymbol p}_j[i] + \hat{\boldsymbol
Q}[i] \hat{\boldsymbol f}_{j,l}[i-1]), \label{adff}
\end{equation}
Once the feedforward filters are computed the feedback filters can
be obtained by
\begin{equation}
\hat{\boldsymbol f}_{j,l} [i] =  \beta_{j,l} {\boldsymbol \Pi}_{j,l}
\hat{\boldsymbol Q}^H[i] \hat{\boldsymbol w}_{j,l}[i]. \label{adfbf}
\end{equation}
Note that the filters need to be initialized and that the
computation of $\hat{\boldsymbol R}^{-1}[i]$ is common to all
branches, i.e., we only need to compute it once for all branches. A
summary of the adaptive MB-MMSE-DF detector is given in Table II.
The receive filters are computed in an alternating fashion, i.e.,
one receive filter is updated followed by the other and the cycle is
repeated for every data symbol. Note that the RLS-type algorithm
presented in Table 1 is a standard version that might need
modifications for a numerically-stable hardware implementation. In
the case of a hardware implementation, a square root (or QR
decomposition) version will have better numerical properties because
the square-root structures do not amplify numerical errors and tend
to assume values within smaller dynamic ranges \cite{haykin}. Other
advanced algorithms might also be considered
\cite{jio_spl}-\cite{delamaretvt}.

\begin{table}[h]
\centering%
\caption{ Proposed Adaptive MB-MMSE-DF detection algorithm.} {
\begin{tabular}{l}
\hline  1. Initialize parameters: ordering, $L$, ${\boldsymbol
S}_{j,l}$, $\beta_{j,l}$, $N_T$ and
$N_R$. \\
~~~~For ~$i=1, ~\ldots,~ Q$, ~where $Q$ is the packet size do \\
2. Compute $\hat{\boldsymbol R}^{-1}[i]$ as follows \\
~~~~${\boldsymbol k}[i] = \frac{\lambda^{-1} \hat{\boldsymbol
R}[i-1] {\boldsymbol r}[i] }{1+ \lambda^{-1} {\boldsymbol r}^H[i]
\hat{\boldsymbol R}[i-1] {\boldsymbol r}[i]},$ \\
~~~~$\hat{\boldsymbol R}[i] = \lambda^{-1} \hat{\boldsymbol R}[i-1]
- \lambda^{-1} {\boldsymbol k}[i] {\boldsymbol r}^H[i] {\boldsymbol
R}^{-1}[i-1]$.
\\ 3. Obtain $\hat{\boldsymbol Q}[i]$ as given by \\
~~~~$\hat{\boldsymbol Q}[i] = \lambda \hat{\boldsymbol Q}[i-1] +
{\boldsymbol r}[i] \hat{\boldsymbol s}_{l_{j,{\rm opt}}}^H[i].$\\
4. Compute $\hat{\boldsymbol p}_j[i]$ for $j=1, \ldots, L$
as follows \\
~~~~$\hat{\boldsymbol p}_j[i] = \lambda \hat{\boldsymbol p}[i-1] +
{\boldsymbol r}[i]s_{j,l_{\rm opt}}^*[i].$\\
5. Determine the ordering $o_{1,l}, \ldots, o_{N_T,l}$ ~for ~ $l=1,
~\ldots,~ L$ \\
6. For ~$l=1, ~\ldots,~ L$~ and ~ $j=1,~\ldots,~N_T$~ compute \\
~~~~~ $\hat{\boldsymbol w}_{j,l}[i] = \hat{\boldsymbol R}^{-1}[i]
\hat{\boldsymbol H}[i] ({\boldsymbol \delta}_j + \hat{\boldsymbol
f}_{j,l}[i-1])$\\
~~~~~ $\hat{\boldsymbol f}_{j,l} [i] =  \beta_{j,l} {\boldsymbol
\Pi}_{j,l} \hat{\boldsymbol H}^H[i] \hat{\boldsymbol w}_{j,l}[i]$ \\
7. For ~$l=1, ~\ldots,~ L$~ and ~$j=1,~\ldots,~N_T$ do \\
~~~~~ Obtain ${ z}_{j,l}[i] = \hat{\boldsymbol
w}^{H}_{j,l}[i]{\boldsymbol r}[i] - \hat{\boldsymbol f}_{j,l}^{H}[i]
\hat{\boldsymbol s}_{l}[i]$ \\
~~~~~ Determine $l_{j,{\rm opt}} = \arg \min_{1 \leq l_j \leq L}
{\rm IMMSE}
(s_j[i],\hat{\boldsymbol w}_{j,l}[i], \hat{\boldsymbol f}_{j,l}[i])$ \\
~~~~~ Detect symbol: $\hat{s}_{j}[i]  = Q \big[ {\boldsymbol z}_{j,l_{\rm opt}}[i] \big] $ \\
 \hline
\end{tabular}
}
\end{table}

\section{Iterative Soft-Input Soft-Output Detection and Decoding}

 {This section presents an iterative version of the
proposed MB-MMSE-DF detector operating with soft-input soft-output
detection and decoding, and with convolutional codes
\cite{wang}-\cite{choi}.} The motivation for the proposed scheme is
that significant gains can be obtained from iterative techniques
with soft cancellation methods and channel codes
\cite{wang}-\cite{choi} when combined with efficient receiver
algorithms. A low-complexity iterative MB-MMSE-DF receiver that
works with a reduced list of candidate symbols and log-likelihood
ratios (LLRs), and that can approach the performance of the optimal
detector is developed. The MIMO system described in Section II is
considered with convolutional codes. The proposed iterative receiver
structure consists of the following stages: a soft-input-soft-output
(SISO) MB-MMSE-DF detector and a maximum \textit{a posteriori} (MAP)
decoder.  {The receiver structure also incorporates a selection
strategy for the list of LLRs which are used to refine the exchange
of soft information.} These stages are separated by interleavers and
deinterleavers. The soft outputs from the MB-MMSE-DF are used to
estimate LLRs which are interleaved and serve as input to the MAP
decoder for the convolutional code. The MAP decoder computes
\textit{a posteriori} probabilities (APPs) for each stream's encoded
symbols, which are used to generate soft estimates. These soft
estimates are subsequently used to update the receive filters of the
MB-MMSE-DF detector, de-interleaved and fed back through the
feedback filter. The MB-MMSE-DF detector computes the \textit{a
posteriori} log-likelihood ratio (LLR) of a transmitted symbol ($+1$
or $-1$) for every code bit of each data stream as given by
 {
\begin{equation}
\begin{split}
\Lambda_1[b_{j,c,l}[i]] & = {\rm log} \frac{P[b_{j,c,l}[i]=+1|{\bf
r}[i]]}{P[b_{j,c,l}[i]=-1| {\bf r}[i]]},  ~j=1, \ldots,  N_T,  ~c=1,
\ldots,  C,  ~l=1, \ldots,L,
\end{split}
\end{equation} }
where $C$ is the number of bits used to map the constellation. Using
Bayes' rule, the above equation can be written as
\begin{equation}
\begin{split}
\Lambda_1[b_{j,c,l}[i]] & = {\rm log} \frac{P[{\bf
r}[i]|b_{j,c,l}[i]=+1]}{P[ {\bf r}[i]|b_{j,c,l}[i]=-1]} + {\rm log}
\frac{P[b_{j,c}[i]=+1]}{P[b_{j,c}[i]=-1]}  = \lambda_1[b_{j,c,l}[i]]
+ \lambda_2^p[b_{j,c}[i]], \label{llr}
\end{split}
\end{equation}
where $\lambda_2^p[b_{j,c}[i]] = {\rm log}
\frac{P[b_{j,c}[i]=+1]}{P[b_{j,c}[i]=-1]}$ is the \textit{a priori} LLR
of the code bit $b_{j,c}[i]$, which is computed by the MAP decoder
processing the $j$th data stream in the previous iteration,
interleaved and then fed back to the MB-MMSE-DF detector. The
superscript $^p$ denotes the quantity obtained in the previous
iteration. Assuming equally likely bits, we have
$\lambda_2^p[b_{j,c}[i]] =0$ in the first iteration for all streams.
The quantity $\lambda_1[b_{j,c,l}[i]] = {\rm log} \frac{P[{\bf
r}[i]|b_{j,c,l}[i]=+1]}{P[ {\bf r}[i]|b_{j,c,l}[i]=-1]}$ represents the
\textit{extrinsic} information computed by the SISO MB-MMSE-DF
detector based on the received data ${\bf r}[i]$, and the prior
information about the code bits $\lambda_2^p[b_{j,c}[i]],~ j=1,\ldots,
N_T,~ c=1,\ldots,
C$  and the $i$th data symbol. Unlike prior work on soft
interference cancellation \cite{wang,lee,choi} and list sphere
decoders \cite{guo,shim,hochwald}, the extrinsic information
$\lambda_1[b_{j,c,l}[i]]$ is obtained from a list of candidate symbols
generated by the MB-MMSE-DF detector and the prior information
provided by the MAP decoder, which is de-interleaved and fed back
into the MAP decoder of the $j$th data stream as the \textit{a
priori} information in the next iteration.

For the MAP decoding, we assume that the interference plus noise at
the output ${\boldsymbol z}_{j,l}[i]$ of the linear receive filters
is Gaussian. This assumption has been reported in previous works
\cite{wang}-\cite{choi} and provides an efficient and accurate way
of computing the extrinsic information. Thus, for the $j$th stream,
the $l$th branch and the $q$th iteration the soft output of the
MB-MMSE-DF detector is
\begin{equation}
z_{j,l}^{(q)}[i] = V_{j,l}^{(q)} s_{j,l}[i] + \xi_{j,l}^{(q)}[i],
\label{output}
\end{equation}
where $V_{j,l}^{(q)}[i]$ is a scalar variable equivalent to the
magnitude of the channel corresponding to the $j$th data stream and
$\xi_{j,l}^{(q)}[i]$ is a Gaussian random variable with variance
$\sigma^2_{\xi_{j,l}^{(q)}}$. Since we have
\begin{equation}
V_{j,l}^{(q)}[i] = E\big[ s_{j,l}^*[i] z_{j,l}^{(q)}[i] \big]
\end{equation}
and
\begin{equation}
\sigma^2_{\xi_{j,l}^{(q)}}[i]  = E\big[ | z_{j,l}^{(q)}[i] -
V_{j,l}^{(q)}[i] s_{j,l}[i]|^2 \big],
\end{equation}
the receiver can obtain the estimates ${\hat V}_{j,l}^{(q)}[i]$ and
${\hat \sigma}^2_{\xi_{j,l}^{(q)}}[i]$ via corresponding sample
averages {over the received symbols}. These estimates are used to
compute the \textit{a posteriori} probabilities $P[b_{j,c,l}[i] =
\pm 1 | z_{j,l}^{(q)}[i]]$ which are de-interleaved and used as
input to the MAP decoder. In what follows, it is assumed that the
MAP decoder generates APPs $P[b_{j,c,l}[i] = \pm 1]$, which are used
to compute the input to the feedback filter ${\boldsymbol f}_{j,l}$.
From (\ref{output}) the extrinsic information generated by the
iterative MB-MMSE-DF is given by
\begin{equation}
\begin{split}
\lambda_1[b_{j,c,l}[i]] & = {\rm log}
\frac{P[z_{j,l}^{(q)}(i)|b_{j,c,l}[i]=+1]}{
P[z_{j,l}^{(q)}[i]|b_{j,c,l}[i]=-1] }  = \log
\frac{\sum\limits_{{\mathbb S} \in {\mathbb S}_c^{+1}} \exp \Big(
-\frac{|z_{j,l}^{(q)}[i] - V_{j,l}^{(q)}{\mathbb S}|^2}{ 2
\sigma^2_{\xi_{j,l}^{(q)}}[i]} \Big)} {\sum\limits_{{\mathbb S} \in
{\mathbb S}_c^{-1}} \exp \Big( -\frac{|z_{j,l}^{(q)}[i] -
V_{j,l}^{(q)}{\mathbb S}|^2}{ 2 \sigma^2_{\xi_{j,l}^{(q)}}[i]}
\Big)},
\end{split}
\end{equation}
where ${\mathbb S}_c^{+1}$ and ${\mathbb S}_c^{-1}$ are the sets of
all possible constellations that a symbol can take on such that the
$c$th bit is $1$ and $-1$, respectively. {Different approaches are
possible to compute the extrinsic information generated from the
list of soft estimates provided by the iterative MB-MMSE-DF
detector. In this work, the iterative MB-MMSE-DF detector chooses
the LLR from a list of $L$ candidates for the decoding iteration as}
\begin{equation}
\lambda_1[b_{j,c,l_{\rm opt}}[i]]  = \arg \max_{1 \leq l \leq L}
\lambda_1[b_{j,c,l}[i]]
\end{equation}
where the selected estimate is the value $\lambda_1[b_{j,c,l_{\rm
opt}}[i]]$ which maximizes the likelihood and corresponds to the
most likely bit. Based on the selected prior information
$\lambda_1^p[b_{j,c,l_{\rm opt}}[i]]$ and the trellis structure of
the code, the MAP decoder processing the $j$th data stream and the
$l$th branch computes the \textit{a posteriori} LLR of each coded
bit as described by
\begin{equation}
\begin{split}
\Lambda_2[b_{j,c}[i]] & = {\rm log} \frac{P[b_{j,c}[i]=+1|
\lambda_1^p[b_{j,c,l_{\rm opt}}[i]; {\rm decoding}]}{P[b_{j,c}[i]=-1|
\lambda_1^p[b_{j,c,l_{\rm opt}}[i]; {\rm decoding}]}
\\ & = \lambda_2[b_{j,c}[i]] + \lambda_1^p[b_{j,c,l_{\rm opt}}[i]], ~ {\rm for} \; j=1, \ldots,
N_T, ~c=1, \dots, C.
\end{split}
\end{equation}
 {The computational burden can be significantly
reduced using the max-log approximation}. From the above, it can be
seen that the output of the MAP decoder is the sum of the prior
information $\lambda_1^p[b_{j,c,l_{\rm opt}}[i]]$ and the extrinsic
information $\lambda_2[b_{j,c}[i]]$ produced by the MAP decoder.
This extrinsic information is the information about the coded bit
$b_{j,c}[i]$ obtained from the selected prior information about the
other coded bits $\lambda_1^p[b_{j,c,l_{\rm opt}}[k]], ~ j \neq i$
\cite{wang}. The MAP decoder also computes the \textit{a posteriori}
LLR of every information bit, which is used to make a decision on
the decoded bit at the last iteration. After interleaving, the
extrinsic information obtained by the MAP decoder
$\lambda_2[b_{j,c}[i]]$ for $j=1, \ldots N_T$, $c=1, \dots, C$ is
fed back to the MB-MMSE-DF detector, as the prior information about
the coded bits of all streams in the subsequent iteration.
 {For the first iteration, $\lambda_1[b_{j,c}[i]]$ and
$\lambda_2[b_{j,c,l_{\rm opt}}[i]]$ are statistically independent
and as the iterations are computed they become more correlated and
the improvement due to each iteration is gradually reduced. A study
of the proposed iterative MB-MMSE-DF detector has indicated that
there is no performance gain when using more than $5$ iterations.}

\section{Analysis of the MB-MMSE-DF Algorithm}

In this section, the computational complexity required by the
MB-MMSE-DF algorithm is evaluated and the diversity order achieved
by the proposed MB-MMSE-DF detector is discussed.

\subsection{Computational Complexity}

The computational complexity of the MB-MMSE-DF detector can be
exactly computed as a function of the number of receive antennas
$N_R$, transmit antennas $N_T$ and branches $L$, as depicted in
Table III. This is in contrast to the SD and the LR-aided
techniques, which require the use of bounds or the counting of
floating point operations (flops). In this study of the
computational cost of the MB-MMSE-DF and other techniques, two
approaches to assess the complexity are employed, namely, the number
of arithmetic operations such as multiplications and additions, and
the number of flops computed by the Lightspeed toolbox \cite{minka}.
\footnote{According to the Lightspeed toolbox \cite{minka} the
number of flops count as $2$ for a complex addition and as $6$ for a
complex multiplication.} The complexity of the SD is associated with
$M(\cdot)$, the Gamma function $\Gamma(\cdot)$, { and the
k-dimensional sphere radius $d_SD$, which is chosen as} a scaled
version of the variance of the noise \cite{hassibi}. The channel
estimation with an RLS algorithm requires $N_RN_T^2 + 4N_T^2 +
2N_TN_R + 2N_T+2$ multiplications and $N_RN_T^2 + 4N_T^2- N_T$
additions.

\begin{table}[h]
\centering%
\caption{\small Computational complexity of detection algorithms per
received vector.} {
\begin{tabular}{ccc}
\hline \rule{0cm}{1.5ex}&  \multicolumn{2}{c}{\small Number of
operations per symbol } \\  \cline{2-3} {\small Algorithm} & {\small
Additions} & {\small Multiplications} \\ \hline
\emph{ \bf  } & {\small $2N_R^2 + N_RN_T -1$} & {\small $3N_R^2+2N_RN_T $}   \\
\emph{\small \bf MB-MMSE-DF + RLS} & {\small $ + L(3N_RN_T^2+2N_T^2 $} & {\small $ +3N_R+1$}   \\
\emph{\bf } & {\small $ -3N_RN_T+N_R-N_T$} & {\small $ +
L(5N_RN_T^2+2N_R)$}
\\ \\
\emph{\small \bf SIC + RLS \cite{rontogiannis}}  & {\small $
\frac{2}{3} N_R^3+\frac{11}{2}N_R^2+4N_R$} & {\small $\frac{2}{3}
N_R^3+\frac{25}{2}N_R^2+3N_R$}
\\ \\
\emph{\small \bf Linear + RLS}  & {\small $N_T(3N_R^2+2N_R - 1) $} & {\small $N_T(3N_R^2+4N_R+1) $} \\
\emph{ }  & {\small $ + 2N_RN_T$} & {\small $$} \\
\\
\emph{\small \bf SD \cite{hassibi}}  & {\small $\sum_{k=1}^{N_T}
\frac{M(k+1)\pi^{k/2}}{\Gamma(k/2+1)} d^k_{\rm SD} $}
& {\small $\sum_{k=1}^{N_T}\frac{M(k) \pi^{k/2}}{\Gamma(k)/2+1) } d^k_{\rm SD} $} \\
\emph{\small \bf  }  & {\small $+ 2N_T^2 - N_T +2 $}
& {\small $+ 2N_T^2$} \\
\hline
\end{tabular}
}
\end{table}

\begin{figure}[!htb]
\begin{center}
\def\epsfsize#1#2{1\columnwidth}
\epsfbox{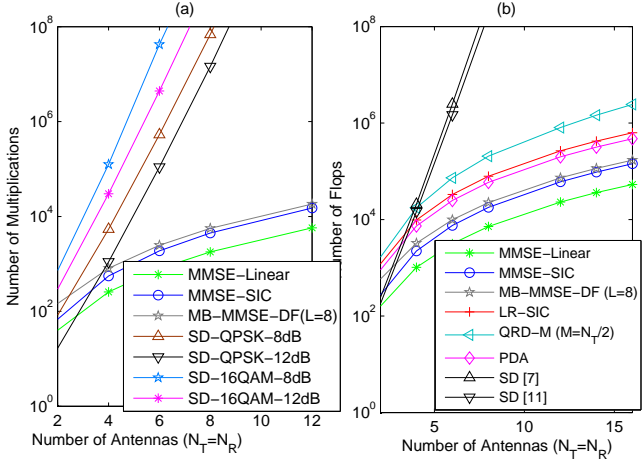} \vspace*{-0.5em} \caption{Computational
complexity of detection algorithms.} \label{complexity}
\end{center}
\end{figure}

An example of the computational cost of some detection algorithms is
shown in Fig. \ref{complexity}, where the number of multiplications
and flops per received vector ${\boldsymbol r}[i]$ are shown for the
proposed MB-MMSE-DF and RLS algorithms, the bounds on the SD
reported by \cite{hassibi} and the SD schemes of
\cite{damen},\cite{shim}, the complex LR-SIC of \cite{gan} with an
MMSE filter, the PDA algorithm reported in \cite{shang} with $I =5$
iterations, the linear detector and the SIC detector
\cite{rontogiannis}. The complexity evaluated in terms of flops
assumes $16$-QAM modulation and ${\rm SNR} = 8$ dB  { and includes
the QRD-M detector \cite{kim_qrdm,kim_qrdm2} with $M=8$. The QRD-M
algorithm is a breadth-first tree search algorithm, whereas the LSD
is a depth-first tree search algorithm that can achieve the optimal
performance. Differently from the QRD-M algorithm and the LSD
algorithms, the proposed MB-MMSE-DF detector associates branches
with different orderings and pairs of linear and feedback filters
that only require one matrix inversion. The list of candidates in
the MB-MMSE-DF algorithm is different because the candidates are
generated by MMSE filtering and feedback cancellation with different
orderings, while the list is generated from a tree search in the
case of the QRD-M detector and the LSD algorithm. Moreover, the
complexity of the MB-MMSE-DF detector only depends on the number of
branches regardless of the constellation size and the
signal-to-noise ratio (SNR), whereas the complexity of the QRD-M
depends on the choice of $M$ and the cost of LSD algorithms is
dependent on the constellation size and the SNR. The curves of Fig.
\ref{complexity} indicate that the proposed MB-MMSE-DF and RLS
algorithms have a complexity higher than the SIC \cite{rontogiannis}
and significantly lower than the SD algorithms for $N_R=N_T > 4$ and
the QRD-M algorithm}. The MB-MMSE-DF detector also has a lower
complexity than the LR-SIC and the PDA algorithms.

\subsection{Diversity Order}

 { The aim of this part is to examine the diversity
order achieved by the MB-MMSE-DF detector. In the analysis, it is
assumed that the data transmission is over a block fading channel,
there is no error propagation due to interference cancellation and
that the SNR is sufficiently high \cite{tse,hughes} (in this case
the MMSE and zero-forcing receive filters have a similar behavior).
The diversity order \cite{tse,hughes} is defined by
\begin{equation}
d \triangleq  \lim_{\textrm{SNR} \rightarrow \infty} \frac{\log
P_{e}(\textrm{SNR})}{\log(\textrm{SNR})},
\end{equation}
where $P_{e}$ denotes the probability of error and  { $\textrm{SNR}
= 10 \log_{10} \frac{N_T \sigma_s^2}{R C ~\sigma^2}$} is the
signal-to-noise ratio, $R$ is the rate of the code and $C$ is the
number of bits required to map the constellation points. It is known
that the diversity order is $d=N_R$ for ML receivers and
$d=N_R-N_T+1$ for receivers with  {SIC} \cite{tse,hughes}. Since for
non-ergodic scenarios the error probability is dominated by the
outage probability \cite{tse,hughes}, the diversity order can be
expressed as
\begin{equation}
d \triangleq  \lim_{x \rightarrow \infty} \frac{ \log Pr( R_{j, {\rm
span}\{1,2,\ldots,j-1,j+1,\ldots, N_T\}} \leq x )}{ \log(x)},
\end{equation}
where $R_{j,{\rm span}\{1,2,\ldots,j-1,j+1,\ldots, N_T\}} =
R_{j,{\rm span}\{\bar{j}\}}$ is the squared projection height from
the $j$th column vector ${\boldsymbol h}_j$ of ${\boldsymbol H}$,
i.e., $R_{j,{\rm span} \{ \bar{j} \}} = ||{\boldsymbol P}
{\boldsymbol h}_j||^2$, where ${\boldsymbol P} = {\boldsymbol I}
-{\boldsymbol B}{\boldsymbol B}^H$ is the projection matrix onto the
orthogonal space of ${\rm span}\{\bar{j}\}$, and ${\boldsymbol B}$
is composed of any orthogonal basis of this subspace.}

 { \textit{Theorem: The diversity order achieved by
the MB-MMSE-DF detector is given by
\begin{equation}
\begin{split}
d_{\rm MB} & \triangleq \lim_{x \rightarrow 0} \frac{ \log Pr(
R_{j,l_{j,{\rm opt}}, {\rm span}\{\bar{l}_{j,{\rm opt}}\}} \leq x
)}{ \log(x)}  = N_R-N_T + G,
\end{split}
\end{equation}
where $1\leq G\leq N_T$ is the number of interference free
candidates among the $L$ candidates for each stream.} }

 { \textit{Proof:} In order to prove this theorem, it
is necessary to make a few assumptions that are common to works that
analyze the diversity order of detectors. The approach used to prove
the theorem is based on induction and the inclusion of an increasing
number of branches that correspond to extra degrees of freedom.}

 { The first assumption is that for each data stream
and branch there is  {an associated diversity order given by} $d
\triangleq \lim_{x \rightarrow \infty} \frac{ \log Pr( R_{j,l,{\rm
span}\{\bar{j}\}} \leq x )}{ \log(x)}$, as established in
\cite{tse,hughes} for a conventional receiver performing SIC in a
MIMO system with $N_T$ transmit and $N_R$ receive antennas. Another
assumption is that the ordering algorithm can exploit the multiple
branches to move each data stream to the last position in the
sequence of detection to obtain an interference free candidate for
detection.}

 { Starting from this point, the result can be
extended by induction. By gradually adding branches with different
orderings, the number of detection candidates available can be
represented by the following sets
\begin{equation}
\begin{split}
{\mathcal S}_{1} & = \{ R_{j,1,{\rm span}\{\bar{j}\}} \},  \\
{\mathcal S}_{2} & = \{ R_{j,1,{\rm span}\{\bar{j}\}}, R_{j,2,{\rm
span}\{\bar{j}\}} \},\\
\vdots &  \\
{\mathcal S}_{L} & = \{ R_{j,1,{\rm span}\{\bar{j}\}}, R_{j,2,{\rm
span}\{\bar{j}\}}, \ldots, R_{j,L,{\rm span}\{\bar{j}\}} \}.\\
\end{split}
\end{equation}
Unlike a conventional receiver with SIC, the proposed MB-MMSE
detector has at any given stage $L$ alternatives to select the
candidate for detection. In fact, the detection of each stream
involves the selection of the best out of $L$ candidate symbols
using the rule in (\ref{eq:error}). The number of degrees of freedom
will depend on the quantities $R_{j,l_{j,{\rm opt}}, {\rm
span}\{\bar{l}_{j,{\rm opt}}\}}$ and wether they correspond to
interference free candidates. }

 {By defining $1\leq G \leq N_T$ as the number of
interference free candidates and assuming that $L> G$ is
sufficiently large to provide a sufficiently large number of
interference free candidates for the $j$th stream, the diversity
order associated with each of the above sets can be described by
\begin{equation}
\begin{split}
 d_{\rm MB}({\mathcal S}_{1}) & = \lim_{x \rightarrow 0}
\frac{ \log Pr( R_{j,l_{j,{\rm opt}}, {\rm span}\{\bar{l}_{j,{\rm
opt}}\}} \leq x )}{
\log(x)}  = N_R-N_T + 1  , \\
N_R-N_T + 1 \leq d_{\rm MB}({\mathcal S}_{2}) & = \lim_{x
\rightarrow 0} \frac{ \log Pr( R_{j,l_{j,{\rm opt}}, {\rm
span}\{\bar{l}_{j,{\rm opt}}\}} \leq x )}{
\log(x)} \leq N_R-N_T + 2  , \\
\vdots & \\
N_R-N_T + 1 \leq d_{\rm MB}({\mathcal S}_{L}) & = \lim_{x
\rightarrow 0} \frac{ \log Pr( R_{j,l_{j,{\rm opt}}, {\rm
span}\{\bar{l}_{j,{\rm opt}}\}} \leq x )}{ \log(x)} \leq N_R-N_T + G
.
\end{split}
\end{equation}
where
\begin{equation}
l_{j,{\rm opt}} = \arg \min_{1\leq l_j \leq L} {\rm IMMSE}
(\sigma_s^2, {\boldsymbol w}_{j,l}, {\boldsymbol f}_{j,l})
\end{equation}
and $R_{j,l_{j,{\rm opt}}, {\rm span}\{\bar{l}_{j,{\rm opt}}\}}$ is
the squared projection height resulting from the selection of the
best out of the available candidates from the set ${\mathcal S}_{l}$
for the $j$th stream. If interference free candidates are gradually
included in the sets and are selected by the above procedure, then
the MB-MMSE-DF detector can obtain $G$ interference free candidates
resulting from $N_T -1$ cancellations for any branch. Hence, the
diversity order for each stream of the MB-MMSE-DF detector is given
by
\begin{equation}
\begin{split}
d_{\rm MB} & \triangleq \lim_{x \rightarrow 0} \frac{ \log Pr(
R_{j,l_{j,{\rm opt}}, {\rm span}\{\bar{l}_{j,{\rm opt}}\}} \leq x
)}{ \log(x)}  = N_R-N_T + G.
\end{split}
\end{equation}
This suggests that the key advantage of the MB-MMSE-DF detector is
its ability to generate $L$ candidates for each stream and select
$G$ interference free candidates. In practice, $G$ will depend on
the number of branches used, the ordering algorithm and the accuracy
of the interference cancellation. }

\section{Simulations}

In this section, the bit error ratio~(BER) performance of the
MB-MMSE-DF and other relevant MIMO detection schemes is evaluated.
 {The sphere decoder (SD) \cite{shim}, the linear
\cite{duel_mimo}, the SIC \cite{vblast}, the QRD-M
\cite{kim_qrdm,kim_qrdm2} with $M=8$, the PDA \cite{jia,lee} with $I
=5$ iterations, MMSE estimators and the proposed MB-MMSE-DF
techniques without and with error propagation mitigation techniques
are considered in the simulations}. The lattice-reduction aided
versions of the linear and the SIC detectors \cite{gan}, which are
denoted LR-MMSE-Linear and LR-MMSE-SIC, respectively, are also
included in the study. The channel coefficients are either static
and obtained from complex Gaussian random variables with zero mean
and unit variance, or time-varying with the coefficients given by
the Jakes model \cite{rappa}. The modulation employed is either QPSK
or $16$-{\rm QAM}. Both uncoded and coded systems are considered.
For the coded systems and iterative detection and decoding, a
non-recursive convolutional code with rate $R=1/2$, constraint
length $3$, generator polynomial $g = [ 7~ 5 ]_{\rm oct}$ and
 { $5$ decoding iterations is adopted}. The numerical
results are averaged over $10^6$ runs, packets with $Q=500$ symbols
for uncoded systems and $Q=1000$ coded symbols are employed and the
signal-to-noise ratio (SNR) in dB is defined as $\textrm{SNR} = 10
\log_{10} \frac{N_T \sigma_s^2}{R
 C ~\sigma^2}$, where $\sigma_s^2$ is the variance of the
symbols, $\sigma^2$ is the noise variance, $R<1$ is the rate of the
channel code and $C$ is the number of bits used to represent the
constellation.

\begin{figure}[!htb]
\begin{center}
\def\epsfsize#1#2{1\columnwidth}
\epsfbox{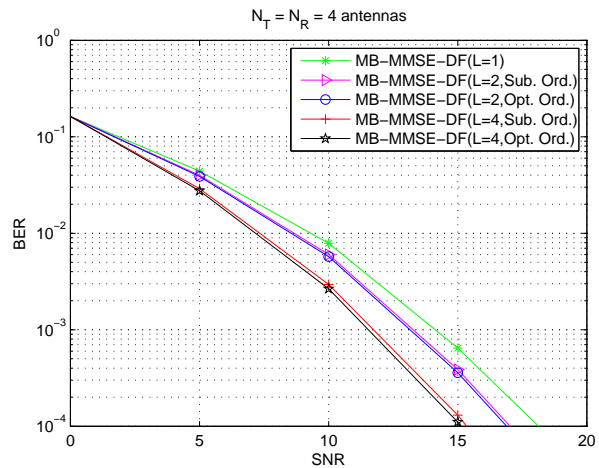} \vspace*{-0.5em} \caption{\small BER performance
of the optimal and the proposed suboptimal ordering algorithms .}
\label{ber_ord}
\end{center}
\end{figure}

{In the first example, the ordering algorithms described in Section
IV are assessed with the MB-MMSE-DF detector using $L=1, 2, {\rm
and}~4$ branches. A MIMO system with $N_T=N_R=4$ antennas is
considered with perfect channel estimation. The BER performance of
the MB-MMSE-DF detector is evaluated with the optimal and the
suboptimal ordering algorithms and the curves are shown in Fig.
\ref{ber_ord}. The results show that the suboptimal ordering
algorithm is able to approach the performance of the optimal
ordering algorithm that performs an exhaustive search. In
particular, the BER performance gap between the optimal and
suboptimal ordering algorithms is small and this has also been
observed for larger systems and a different number of branches $L$.
For this reason, the suboptimal ordering algorithm has been adopted
in the next examples.}

\begin{figure}[!htb]
\begin{center}
\def\epsfsize#1#2{1\columnwidth}
\epsfbox{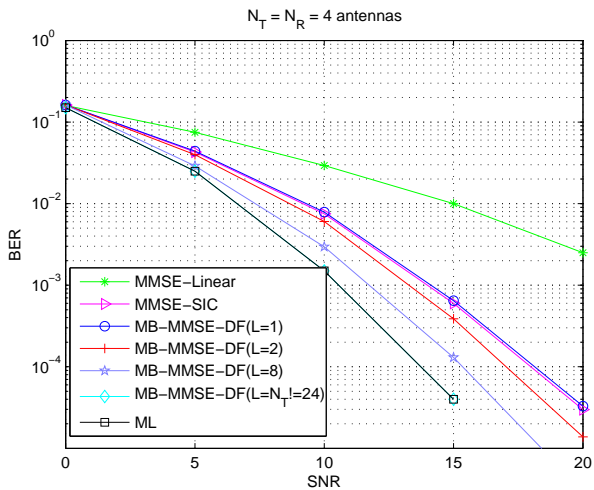} \vspace*{-0.5em} \caption{\small BER performance
of the detectors with perfect channel estimation for multiple
branches $L$ and QPSK.} \label{ber_difL}
\end{center}
\end{figure}

The uncoded BER performance of the proposed MB-MMSE-DF detector is
then considered in an example to evaluate the number of branches
that should be used in the suboptimal ordering algorithm. It is also
important to account for the impact of additional branches on the
performance with perfect channel estimation for a MIMO system with
$N_T=N_R=4$ antennas. The proposed suboptimal ordering algorithm is
compared against the optimal ordering approach described in Section
IV that evaluates $N_T!=24$ possible branches.  {The MB-MMSE-DF
detector has been designed with $L=2 ~{\rm and}~4$ parallel branches
and its BER performance against the SNR has been compared with those
of the existing schemes, as depicted in Fig. \ref{ber_difL}. In
fact, the MB-MMSE-DF detector is able to gradually approach the BER
performance of the ML detector as the number of branches L is
increased. Starting with L=1, the MB-MMSE-DF detector has a BER
performance comparable with that of a standard MMSE-SIC detector. By
increasing L the BER performance of MB-MMSE-DF gradually improves
and gets within 1.5 dB of SNR for the same BER performance as the ML
detector when L=8. Finally, MB-MMSE-DF obtains a performance that is
comparable (the curves coincide) when L=24. }

\begin{figure}[!htb]
\begin{center}
\def\epsfsize#1#2{1\columnwidth}
\epsfbox{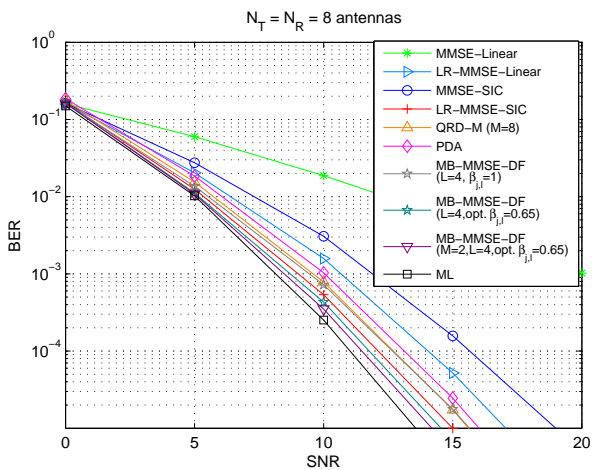} \caption{\small BER performance with adaptive
estimation and QPSK modulation.}\label{ber_snr1}
\end{center}
\end{figure}

\begin{figure}[!htb]
\begin{center}
\def\epsfsize#1#2{1\columnwidth}
\epsfbox{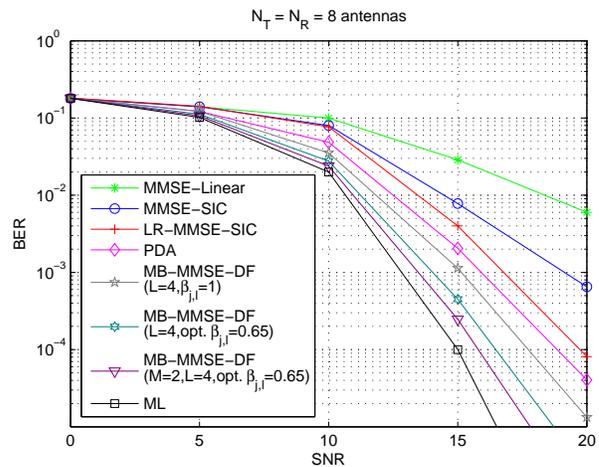} \vspace*{-0.5em} \caption{\small BER performance
with adaptive estimation and $16-${\rm QAM}.} \label{ber_qam}
\end{center}
\end{figure}

In the next experiments depicted in Figs. \ref{ber_snr1} and
\ref{ber_qam}, the uncoded BER performance of the proposed
MB-MMSE-DF detector is evaluated with $L=4$, $N_T=N_R=8$ antennas,
QPSK and $16$-QAM modulation, a block fading channel and adaptive
estimation using the proposed RLS-type algorithm with
$\hat{\boldsymbol R}^{-1}[0]=10^{-2}{\boldsymbol I}$ and $\lambda
=0.998$. In the transmission, we assume packets with $500$ symbols
and employ a training sequence with $N_{Tr} = 50$ symbols to compute
the channel and the receive filter coefficients. We include in the
comparison the linear and SIC detectors with RLS algorithms, the
LR-MMSE-Linear and LR-MMSE-SIC detection schemes \cite{gan} using
MMSE filters, the QRD-M technique of \cite{kim_qrdm,kim_qrdm2}, the
PDA algorithm of \cite{jia} and the SD of \cite{shim} to compute the
ML solution, which employ the RLS algorithm to estimate the
channels.

{The results depicted in Figs. \ref{ber_snr1} and \ref{ber_qam} for
QPSK and $16$-QAM, respectively, show that the proposed MB-MMSE-DF
detector achieves a performance which is close to the ML solutions
implemented with the SD and outperforms the linear, the SIC, the
LR-MMSE-Linear, LR-MMSE-SIC, the QRD-M and the PDA detectors by a
significant margin}.  {In particular, the proposed MB-MMSE-DF
detector with $L=4$ without error propagation mitigation
($\beta_{j,l}= \gamma_{j,l}=1$) has a comparable performance to the
PDA and the LR-MMSE-SIC detectors, whereas the MB-MMSE-DF scheme
with an optimized value of $\beta_{j,l}=0.65$ outperforms the PDA
and the LR-MMSE-SIC schemes. The MB-MMSE-DF scheme with $M=2$ stages
and $L=4$ significantly outperforms the QRD-M, the PDA and the
LR-MMSE-SIC algorithms and achieves a performance within $1$ dB from
the ML solution, while requiring a cost comparable to the SIC with
the RLS algorithm.}

\begin{figure}[!htb]
\begin{center}
\def\epsfsize#1#2{1\columnwidth}
\epsfbox{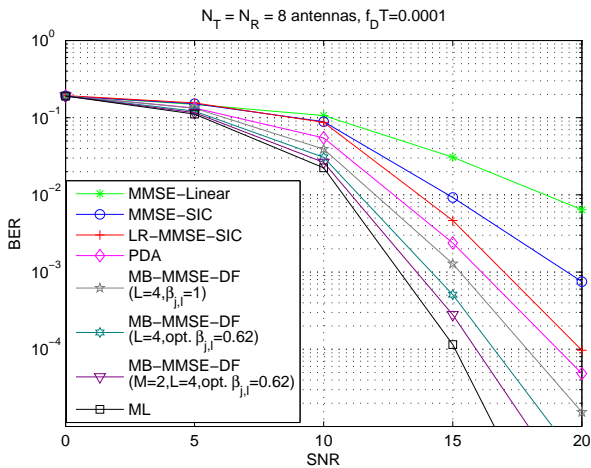} \vspace*{-0.5em} \caption{\small BER performance
with adaptive estimation and $16-${\rm QAM} in time-varying channels
with $f_DT=10^{-4}$.} \label{ber_snr2}
\end{center}
\end{figure}

In the next two examples, the uncoded and coded BER performances of
the detectors are assessed for systems with $16$-QAM modulation and
time-varying channels. The channel coefficients in these examples
change every received vector according to the Jakes model
\cite{rappa} and the results are shown in terms of the normalized
Doppler frequency $f_{D}T$ (cycles/symbol), where $f_D$ is the
maximum Doppler shift and $T$ is the symbol interval. For the data
transmission, packets with $Q=500$ symbols are used for the uncoded
system, with $Q=1000$ coded symbols for the convolutionally coded
system with  {5 decoding iterations}, and a training sequence with
$N_{Tr} = 50$ symbols is employed to compute the channel and receive
filter coefficients. After the training sequence, the receivers are
switched to decision-directed mode and the parameters are tracked
with RLS-type algorithms.

\begin{figure}[!htb]
\begin{center}
\def\epsfsize#1#2{1\columnwidth}
\epsfbox{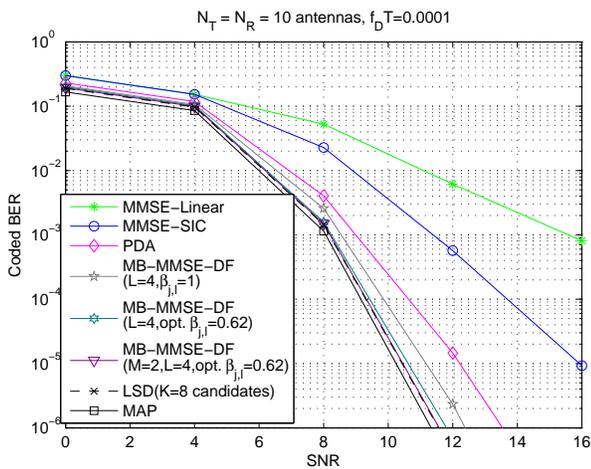} \vspace*{-0.5em} \caption{\small Coded BER
performance with adaptive estimation and $16-${\rm QAM} in
time-varying channels with $f_DT=10^{-4}$ and
 {5 decoding iterations}.}\label{cber}
\end{center}
\end{figure}

{The uncoded BER results illustrated in Fig. \ref{ber_snr2} show
similar results to that of Fig. \ref{ber_qam} but with a slight
performance degradation due to the time-varying nature of the
channel. The coded BER {  performance illustrated in Fig. \ref{cber}
indicates} that the proposed iterative MB-MMSE-DF detector with an
optimized value of $\beta_{j,l}=0.62$, $L=4$ and $M=2$ has a
performance that is very close to the optimal MAP detector and is
comparable to the list SD (LSD) with $K=8$ candidates, which
corresponds to the SD of \cite{shim} with LLR processing.} The
proposed iterative MB-MMSE-DF detector has a gain of up to $2$ dB
over the PDA detector and of up to $5$ dB over the conventional SIC
with iterative processing {  for the same coded BER}, while the
computational cost of MB-MMSE-DF is significantly lower than the PDA
technique and slightly higher than the SIC algorithm.

\begin{figure}[!htb]
\begin{center}
\def\epsfsize#1#2{1\columnwidth}
\epsfbox{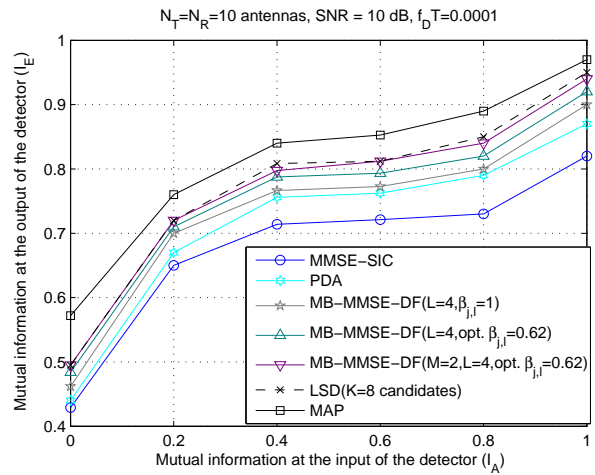} \vspace*{-0.5em} \caption{\small EXIT chart for
the analyzed detectors with $16-${\rm QAM}.} \label{exit}
\end{center}
\end{figure}

In Fig. \ref{exit}, the soft input and output behavior of the
detection algorithms is described through the use of the extrinsic
information transfer (EXIT) chart \cite{tenbrink} analysis. In this
plot, $16$-QAM modulation with a $10 \times 10$ MIMO system are
considered. The quantities $I_A$ and $I_E$ represent the mutual
information at the input and at the output of the detectors
analyzed. The proposed MB-MMSE-DF detector is able to achieve a
higher capacity compared to the other suboptimal algorithms
considered and to follow closely the trajectory of the LSD and MAP
algorithms. Specifically, with the increased number of branches,
more tentative decisions or candidates are included in the search
space for the solution and this allows the proposed MB-MMSE-DF
detector to approach the performance of the MAP algorithm.

\section{Concluding Remarks}

This work has proposed and investigated MB-MMSE-DF detection
algorithms for large MIMO systems using spatial multiplexing.
Constrained MMSE filters designed with constraints on the shape and
magnitude of the feedback filters have been presented for the
MB-MMSE-DF detector and it has been shown that the proposed design
does not require a significant additional complexity over the
conventional MMSE-DF detector. Optimal and sub-optimal ordering
algorithms have also been presented for the MB-MMSE-DF detector. An
adaptive version of the MB-MMSE-DF detector has been developed with
an RLS-type algorithm for estimating the parameters of the filters
when the channel is time-varying. A soft-output version of the
MB-MMSE DF detector has also been proposed as a component of an
iterative detection and decoding receiver structure. The results
have shown that the proposed MB-MMSE-DF detector achieves a
performance superior to some existing suboptimal detectors and close
to the ML detector, while requiring significantly lower complexity.

\begin{appendix}

\section{Derivation of MMSE Receive Filters}

{In order to derive the MMSE receive filters resulting from the
constrained optimization problem presented in (\ref{eq:msedfprop}),
the method of Lagrange multipliers is used which results in the
following unconstrained cost function \vspace{-0.25em}
\begin{equation}
\begin{split}
{\mathcal L}({\boldsymbol w}_{j,l},{\boldsymbol f}_{j,l},
{\boldsymbol \alpha}_{j,l}, \mu_{j,l}) & = E[ |s_{j}[i] -
{\boldsymbol w}_{j,l}^H{\boldsymbol r}[i] + {\boldsymbol f}_{j,l}^H
\hat{\boldsymbol s}_{l}[i]|^2]  + 2\Re[ ({\boldsymbol
S}_{j,l}{\boldsymbol f}_{j,l})^H{\boldsymbol \alpha}_{j,l}] \\ &
\quad + 2\Re[({\boldsymbol f}_{j,l}^H{\boldsymbol f}_{j,l} -
\gamma_{j,l} {\boldsymbol f}_{j,l}^{H,~c}{\boldsymbol
f}_{j,l}^{c})^H\mu_{j,l}], ~~j=1, \ldots, N_T,
~l=1,\ldots,L,\label{lag}
\end{split}
\end{equation} \vspace{-0.25em}
where ${\boldsymbol \alpha}_{j,l}$ is an $N_T \times 1$ vector of
Lagrange multipliers and $\mu_{j,l}$ is a scalar Lagrange
multiplier.

By computing the gradient terms of (\ref{lag}) with respect to
${\boldsymbol w}_{j,l}^*$ and equating them to zero, we have
\vspace{-0.25em}
\begin{equation}
\nabla {\mathcal L}({\boldsymbol w}_{j,l},{\boldsymbol f}_{j,l},
{\boldsymbol \alpha}_{j,l}, \mu_{j,l})_{{\boldsymbol w}_{j,l}^*} =
E[ -{\boldsymbol r}[i](s_{j}[i] - {\boldsymbol
w}_{j,l}^H{\boldsymbol r}[i] + {\boldsymbol f}_{j,l}^H
\hat{\boldsymbol s}_{l}[i])^*] = {\boldsymbol 0}
\end{equation} \vspace{-0.25em}
By further manipulating the terms in the above equation, we arrive
at the expression obtained in (\ref{eq:dfeprop1}) \vspace{-0.25em}
\begin{equation}
{\boldsymbol w}_{j,l}^{\rm MMSE} = {\boldsymbol R}^{-1}({\boldsymbol
p}_{j} + {\boldsymbol Q}{\boldsymbol f}_{j,l}), \label{derw}
\end{equation} \vspace{-0.25em}
where ${\boldsymbol R} = E[{\boldsymbol r}[i] {\boldsymbol r}^H[i]]$
is the $N_R \times N_R$ covariance matrix of the received data,
${\boldsymbol p}_j = E[{\boldsymbol r}[i] s_{j}^{*}[i]]$ is the $N_R
\times 1$ cross-correlation vector and ${\boldsymbol Q} =
E[{\boldsymbol r}[i] \hat{\boldsymbol s}_{l}^{H}[i]]$ is a $N_R
\times N_T$ cross-correlation matrix.

By calculating the gradient terms of (\ref{lag}) with respect to
${\boldsymbol f}_{j,l}^*$ and equating them to zero, we have
\begin{equation} \vspace{-0.25em}
\nabla {\mathcal L}({\boldsymbol w}_{j,l},{\boldsymbol f}_{j,l},
{\boldsymbol \alpha}_{j,l}, \mu_{j,l})_{{\boldsymbol f}_{j,l}^*} =
E[ \hat{\boldsymbol s}_{l}[i](s_{j}[i] - {\boldsymbol
w}_{j,l}^H{\boldsymbol r}[i] + {\boldsymbol f}_{j,l}^H
\hat{\boldsymbol s}_{l}[i])^*] + 2{\boldsymbol S}_{j,l}{\boldsymbol
\alpha}_{j,l} + 2{\boldsymbol f}_{j,l}\mu_{j,l} = {\boldsymbol 0}
\end{equation} \vspace{-0.25em}
 {Using the above equation and with further
manipulations, we obtain
\begin{equation} \vspace{-0.25em}
{\boldsymbol f}_{j,l}^{\rm MMSE} = \frac{\beta_{j,l}}{\sigma_s^2}
({\boldsymbol Q}^H{\boldsymbol w}_{j,l} - {\boldsymbol t}_j -
2{\boldsymbol S}_{j,l}^H {\boldsymbol \alpha}_{j,l}). \label{fint}
\end{equation}
where the term $\beta_{j,l} = (1-2 \mu_{j,l})^{-1}$ with the
Lagrange multiplier $\mu_{j,l}$ is responsible for the mitigation of
the error propagation and is a parameter to be adjusted, and
$E[\hat{\boldsymbol s}[i]{\boldsymbol s}^H[i]] =
\sigma_s^2{\boldsymbol I}$ since it is assumed that
$\hat{\boldsymbol s}[i]$ has independent and identically distributed
entries.} The above expression describes the relationship between
the feedback filters ${\boldsymbol f}_{j,l}^{\rm opt}$, the
feedforward filter ${\boldsymbol w}_{j,l}^{\rm opt}$ and the
quantities ${\boldsymbol Q}$, ${\boldsymbol t}_j =
E[\hat{\boldsymbol s}_{l}[i] s_j[i]]$, and the Lagrange multipliers
$\mu_{j,l}$ and ${\boldsymbol \alpha}_{j,l}$.

The expression for the Lagrange multiplier ${\boldsymbol
\alpha}_{j,l}$ can be obtained by computing the gradient terms of
(\ref{lag}) with respect to ${\boldsymbol \alpha}_{j,l}$ and
equating them to zero, which results in \vspace{-0.25em}
\begin{equation}
\nabla {\mathcal L}({\boldsymbol w}_{j,l},{\boldsymbol f}_{j,l},
{\boldsymbol \alpha}_{j,l}, \mu_{j,l})_{{\boldsymbol \alpha}_{j,l}}
= {\boldsymbol S}_{j,l}{\boldsymbol f}_{j,l} = {\boldsymbol 0}
\end{equation} \vspace{-0.25em}
By substituting (\ref{fint}) into the above expression and solving
for ${\boldsymbol \alpha}_{j,l}$, we have \vspace{-0.25em}
\begin{equation} \vspace{-0.25em}
{\boldsymbol \alpha}_{j,l} = ( {\boldsymbol S}_{j,l} {\boldsymbol
S}_{j,l}^H )^{-1} \big(  {\boldsymbol S}_{j,l}^H {\boldsymbol Q}^H
{\boldsymbol w}_{j,l} - {\boldsymbol S}_{j,l}{\boldsymbol t}_j
\big)/2,
\end{equation} \vspace{-0.25em}
 {It turns out that there is no closed form solution
for the term $\beta_{j,l}$, which is a function of the Lagrange
multiplier $\mu_{j,l}$. This happens because its evaluation leads to
a quadratic function of the feedback filter ${\boldsymbol f}_{j,l}$
that is quite involved. For this reason, we employ an approach that
computes $\beta_{j,l}$ numerically.} By inserting the expression for
${\boldsymbol \alpha}_{j,l}$ into (\ref{fint}), we arrive at
\vspace{-0.25em}
\begin{equation}
\hspace{-1.5em}{\boldsymbol f}_{j,l}^{\rm MMSE} =
\frac{\beta_{j,l}}{\sigma_s^2} \Big({\boldsymbol Q}^H{\boldsymbol
w}_{j,l} - {\boldsymbol t}_j - {\boldsymbol S}_{j,l}^H (
{\boldsymbol S}_{j,l} {\boldsymbol S}_{j,l}^H )^{-1} \big(
{\boldsymbol S}_{j,l}^H {\boldsymbol Q}^H {\boldsymbol w}_{j,l} -
{\boldsymbol S}_{j,l}{\boldsymbol t}_j\big) \Big),
 \label{fint2}
\end{equation}
 {where ${\boldsymbol f}_{j,l}^{\rm MMSE}$ is a
function of the Lagrange multiplier $\mu_{j,l}$.} By defining
${\boldsymbol \Pi}_{j,l} = {\boldsymbol I} - {\boldsymbol
S}_{j,l}^H({\boldsymbol S}_{j,l}^H {\boldsymbol
S}_{j,l})^{-1}{\boldsymbol S}_{j,l}$ as a projection matrix that
ensures the shape constraint ${\boldsymbol S}_{j,l}$, the above
expression can be written as (\ref{eq:dfeprop2}) in the compact form
\vspace{-0.25em}
\begin{equation}
\begin{split}
 {\boldsymbol f}_{j,l}^{\rm MMSE} & = \frac{\beta_{j,l}}{\sigma_s^2}
{\boldsymbol \Pi}_{j,l} ({\boldsymbol Q}^H{\boldsymbol w}_{j,l} -
{\boldsymbol t}_{j})  ,\label{fint3}
\end{split}
\end{equation} \vspace{-0.25em}
Now if we substitute the above expression into (\ref{derw}) and
further manipulate the expressions, we obtain \vspace{-0.25em}
\begin{equation} \vspace{-0.25em}
{\boldsymbol w}_{j,l}^{\rm MMSE} = ({\boldsymbol R} - \beta_{j,l}
{\boldsymbol Q} {\boldsymbol \Pi}_{j,l} {\boldsymbol Q}^H)^{-1}\big(
{\boldsymbol p}_{j} - \beta_{j,l} {\boldsymbol Q} {\boldsymbol
\Pi}_{j,l} {\boldsymbol t}_{j} \big). \label{derw2}
\end{equation} \vspace{-0.25em}
Substituting the above expression into (\ref{fint3}), we obtain
\vspace{-0.25em}
\begin{equation}
\begin{split}
\hspace{-1.5em} {\boldsymbol f}_{j,l}^{\rm MMSE} & =
\frac{\beta_{j,l}}{\sigma_s^2} {\boldsymbol \Pi}_{j,l}
\Big[{\boldsymbol Q}^H({\boldsymbol R} - \beta_{j,l} {\boldsymbol Q}
{\boldsymbol \Pi}_{j,l} {\boldsymbol Q}^H)^{-1}\big( {\boldsymbol
p}_{j} - \beta_{j,l} {\boldsymbol Q} {\boldsymbol \Pi}_{j,l}
{\boldsymbol t}_{j}
 \big) - {\boldsymbol t}_{j}\Big]
,\label{fint4}
\end{split}
\end{equation}\vspace{-0.25em}
where the above expressions for the receive filters ${\boldsymbol
w}_{j,l}$ and ${\boldsymbol f}_{j,l}$ only depend on the statistical
quantities ${\boldsymbol R}$, ${\boldsymbol Q}$, ${\boldsymbol t}_j$
and ${\boldsymbol p}_j$ and the parameters $\beta_{j,l}$ and
${\boldsymbol \Pi}_{j,l}$.  {Nevertheless, these expressions have an
inconvenient form for practical use as they require multiple matrix
inversions for the computation of the receiver filters for each data
stream $j$ and branch $l$. To circumvent this drawback, the use of
an alternating strategy with (\ref{derw}) and (\ref{fint2}) is
employed as it allows a designer to compute only one matrix
inversion (${\boldsymbol R}^{-1}$) and all the receive filters with
a reduced number of extra multiplications and additions.}

The expressions obtained so far can be simplified by evaluating some
of the key statistical quantities such as ${\boldsymbol R}$,
${\boldsymbol Q}$, ${\boldsymbol t}_j$ and ${\boldsymbol p}_j$ and
replacing them in the formulas for the receive filters. Using the
fact that the quantity ${\boldsymbol t}_j= {\boldsymbol 0}$ for
interference cancellation, ${\boldsymbol v}_{j,l}= {\boldsymbol 0}$,
and assuming perfect feedback (${\boldsymbol s} = \hat{\boldsymbol
s}$) we have \vspace{-0.25em}
\begin{equation}
\begin{split}
{\boldsymbol R} & = {\sigma_s}^2{\boldsymbol H}{\boldsymbol H}^H +
{\sigma_n^2}{\boldsymbol I}, {\boldsymbol Q}
={\sigma_s}^2{\boldsymbol H}, {\boldsymbol p}_j  =
{\sigma_s}^2{\boldsymbol H} {\boldsymbol \delta}_j, \label{quant}
\end{split}
\end{equation} \vspace{-0.25em}
where ${\boldsymbol \delta}_j = [ \underbrace{0 \ldots 0}_{j-1} ~1~
\underbrace{0 \ldots 0}_{N_T-j-2}]^T$ is a $N_T \times 1$ vector
with a one in the $j$th element and zeros elsewhere. Substituting
${\boldsymbol t}_j= {\boldsymbol 0}$ and (\ref{quant}) into
(\ref{derw}) and (\ref{fint2}) we arrive at (\ref{eq:dfe1}) and
(\ref{eq:dfe2}), respectively. }

\section{MMSE Expressions and Connections with Conventional DF Receivers}
\vspace{-0.5em} The MMSE associated with the filters ${\boldsymbol
w}_{j,l}$ and ${\boldsymbol f}_{j,l}$ and the statistics of the data
symbols $s_j[i]$ is given by \vspace{-0.25em}
\begin{equation}
\begin{split}
\hspace{-0.75em}\underbrace{{\rm MMSE} (s_j[i],{\boldsymbol
w}_{j,l}^{\rm MMSE}, {\boldsymbol f}_{j,l}^{\rm MMSE})}_{{\rm
MMSE}_j} & = E[|s_j[i] - {\boldsymbol w}_{j,l}^H{\boldsymbol r}[i] +
{\boldsymbol f}_{j,l}^H{\boldsymbol s}_{l}[i]|^2]
\\
& = \underbrace{E[|s_j[i]|^2]}_{\sigma_s^2} - {\boldsymbol
w}_{j,l}^H \underbrace{E[{\boldsymbol r}[i]s_j^*[i]]}_{{\boldsymbol
p}_j} -  \underbrace{E[{\boldsymbol r}^H[i]s_j[i]]}_{{\boldsymbol
p}_j^H}{\boldsymbol w}_{j,l}+ {\boldsymbol
w}_{j,l}^H\underbrace{E[{\boldsymbol r}[i]{\boldsymbol
r}^H[i]]}_{\boldsymbol R} {\boldsymbol w}_{j,l} \\ & \quad  -
{\boldsymbol w}_{j,l}^H \underbrace{E[{\boldsymbol r}[i]
{\boldsymbol s}_{l}^{H}[i]]}_{{\boldsymbol Q}}{\boldsymbol f}_{j,l}
- {\boldsymbol f}_{j,l}^H{\boldsymbol Q}^H{\boldsymbol w}_{j,l} +
{\boldsymbol f}_{j,l}^H \underbrace{E[{\boldsymbol s}_{l}
s_j^*[i]]}_{{\boldsymbol t}_j} + {\boldsymbol t}_j^H{\boldsymbol
f}_{j,l}  + {\boldsymbol f}_{j,l}^H \underbrace{E[{\boldsymbol
s}_{l}{\boldsymbol
s}_{l}^{H}]}_{\bf I} {\boldsymbol f}_{j,l} \\
& =  \sigma_s^2 - {\boldsymbol w}_{j,l}^H{\boldsymbol p}_j -
{\boldsymbol p}_j^H{\boldsymbol w}_{j,l} + {\boldsymbol
w}_{j,l}^H{\boldsymbol R} {\boldsymbol w}_{j,l}   - {\boldsymbol
w}_{j,l}^H{\boldsymbol Q}{\boldsymbol f}_{j,l} - {\boldsymbol
f}_{j,l}^H{\boldsymbol Q}^H{\boldsymbol w}_{j,l}
+ {\boldsymbol f}_{j,l}^H{\boldsymbol t}_j  + {\boldsymbol t}_j^H{\boldsymbol f}_{j,l} +{\boldsymbol f}_{j,l}^H{\boldsymbol f}_{j,l}\\
& =  \sigma_s^2 - {\boldsymbol w}_{j,l}^H{\boldsymbol p}_j -
{\boldsymbol p}_j^H{\boldsymbol w}_{j,l} + \underbrace{({\boldsymbol
f}_{j,l}^H
{\boldsymbol Q}^H + {\boldsymbol p}_j^H)}_{{\boldsymbol w}_{j,l}}{\boldsymbol R}^{-1} {\boldsymbol w}_{j,l} \\
& \quad - {\boldsymbol w}_{j,l}^H{\boldsymbol Q}{\boldsymbol
f}_{j,l} - {\boldsymbol f}_{j,l}^H{\boldsymbol Q}^H{\boldsymbol
w}_{j,l} + {\boldsymbol f}_{j,l}^H{\boldsymbol t}_j  + {\boldsymbol
t}_j^H{\boldsymbol f}_{j,l} +{\boldsymbol f}_{j,l}^H{\boldsymbol
f}_{j,l}\\
& = \sigma_s^2 - {\boldsymbol w}_{j,l}^H {\boldsymbol R}
{\boldsymbol w}_{j,l} + {\boldsymbol t}_j^H{\boldsymbol f}_{j,l}
+{\boldsymbol f}_{j,l}^H{\boldsymbol t}_{j}+ {\boldsymbol f}_{j,l}^H
{\boldsymbol f}_{j,l}.  \label{MMSE_long}
\end{split}
\end{equation}\vspace{-0.25em}
By substituting ${\boldsymbol t}_j = {\boldsymbol 0}$, the
quantities in (\ref{quant}) and the expressions in (\ref{derw2}) and
(\ref{fint4}), the MMSE becomes
\begin{equation} \vspace{-0.25em}
\begin{split}
{\rm MMSE}_j & =
 \sigma_s^2 - {\boldsymbol w}_{j,l}^{H, ~{\rm MMSE}} {\boldsymbol R} {\boldsymbol
w}_{j,l}^{\rm MMSE}  + {\boldsymbol f}_{j,l}^{H, ~{\rm MMSE}}
{\boldsymbol f}_{j,l}^{\rm opt} \\
& = \sigma_s^2 - {\boldsymbol p}_{j}^H({\boldsymbol R} - \beta_{j,l}
{\boldsymbol Q} {\boldsymbol \Pi}_{j,l} {\boldsymbol Q}^H)^{-1}
{\boldsymbol R} ({\boldsymbol R} - \beta_{j,l} {\boldsymbol Q}
{\boldsymbol \Pi}_{j,l} {\boldsymbol Q}^H)^{-1} {\boldsymbol p}_{j}
\\ & \quad + \beta_{j,l}^2{\boldsymbol p}_{j}^H ({\boldsymbol R} -
\beta_{j,l}{\boldsymbol Q} {\boldsymbol \Pi}_{j,l} {\boldsymbol
Q}^H)^{-1} {\boldsymbol Q}{\boldsymbol \Pi}_{j,l}^H {\boldsymbol
\Pi}_{j,l} {\boldsymbol Q}^H({\boldsymbol R} -
\beta_{j,l}{\boldsymbol Q} {\boldsymbol \Pi}_{j,l} {\boldsymbol
Q}^H)^{-1}{\boldsymbol p}_{j} . \label{MMSE_long2}
\end{split}
\end{equation} \vspace{-0.25em}
For a given ordering and branch $l$, the sum of the MMSE in
(\ref{MMSE_long2}) over the $N_T$ data streams is equivalent to the
MMSE achieved by a conventional MMSE-DF receiver (C-MMSE-DF) and is
given by
\begin{equation} \vspace{-0.25em}
{\rm SMMSE} = \sum_{j=1}^{N_T} MMSE_j
\end{equation} \vspace{-0.25em}
 {An instantaneous MMSE metric for the selection of
the best branch for each received vector can be obtained by removing
the expected value from the expression in (\ref{MMSE_long}) and
considering each received data vector ${\boldsymbol r}[i]$, which
results in
\begin{equation}
\begin{split}
\hspace{-0.75em}{\rm IMMSE} (s_j[i],{\boldsymbol w}_{j,l},
{\boldsymbol f}_{j,l}, {\boldsymbol r}[i] ) & = |s_j[i] -
{\boldsymbol w}_{j,l}^H{\boldsymbol r}[i] + {\boldsymbol
f}_{j,l}^H{\boldsymbol s}_{l}[i]|^2
\\
& = {|s_j[i]|^2} - {\boldsymbol w}_{j,l}^H \underbrace{{\boldsymbol
r}[i]s_j^*[i]}_{\hat{\boldsymbol p}_j} - \underbrace{{\boldsymbol
r}^H[i]s_j[i]}_{\hat{\boldsymbol p}_j^H}{\boldsymbol w}_{j,l}+
{\boldsymbol w}_{j,l}^H\underbrace{{\boldsymbol r}[i]{\boldsymbol
r}^H[i]}_{\hat{\boldsymbol R}} {\boldsymbol w}_{j,l} \\ & \quad  -
{\boldsymbol w}_{j,l}^H \underbrace{{\boldsymbol r}[i] {\boldsymbol
s}_{l}^{H}[i]}_{\hat{\boldsymbol Q}}{\boldsymbol f}_{j,l} -
{\boldsymbol f}_{j,l}^H\hat{\boldsymbol Q}^H{\boldsymbol w}_{j,l} +
{\boldsymbol f}_{j,l}^H \underbrace{{\boldsymbol s}_{l}
s_j^*[i]}_{\hat{\boldsymbol t}_j} + {\boldsymbol t}_j^H{\boldsymbol
f}_{j,l}  + {\boldsymbol f}_{j,l}^H {\boldsymbol s}_{l}{\boldsymbol
s}_{l}^{H} {\boldsymbol f}_{j,l} . \label{IMMSE_long}
\end{split}
\end{equation}}
 { The expression above suggests that in order to
obtain an instantaneous MMSE metric, the MB-MMSE-DF detector needs
to compute all the terms. However, it is possible to obtain an
effective and yet more efficient expression by inspecting the terms
in the first line of (\ref{MMSE_long2}) and retaining the
corresponding instantaneous values, which results in
\begin{equation}
{\rm IMMSE} (s_j[i],{\boldsymbol w}_{j,l}, {\boldsymbol f}_{j,l},
{\boldsymbol r}[i] ) \approx
 |s_j[i]|^2 - {\boldsymbol w}_{j,l}^{H, ~{\rm MMSE}} \hat{\boldsymbol R} {\boldsymbol
w}_{j,l}^{\rm MMSE}  + {\boldsymbol f}_{j,l}^{H, ~{\rm MMSE}}
{\boldsymbol s}_{l}{\boldsymbol s}_{l}^{H} {\boldsymbol
f}_{j,l}^{\rm MMSE} \label{IMMSE_long2}.
\end{equation}
The expression in (\ref{IMMSE_long2}) has been tested and compared
with (\ref{IMMSE_long}), and the results indicate an equivalent
performance of the two expressions. Due to the smaller number of
terms, the expression in (\ref{IMMSE_long2}) has been adopted for
the operation of the MB-MMSE-DF detector.}

\end{appendix}

\end{document}